\documentclass{article}

\usepackage{arxiv}

\usepackage[utf8]{inputenc} 
\usepackage[T1]{fontenc}    
\usepackage{booktabs}       
\usepackage{amsfonts}       
\usepackage{nicefrac}       
\usepackage{microtype}      
\usepackage{amsmath}
\usepackage{graphicx}
\usepackage{doi}
\usepackage{caption}
\makeatletter
\newcommand{\topcaption}[2][]{%
    \addtocounter{table}{1}%
    \par\noindent\textbf{Table \thetable.} #2%
    \if\relax\detokenize{#1}\relax\else%
        \def\@currentlabel{\thetable}%
        \label{#1}%
    \fi%
    \par%
}
\makeatother

\usepackage{lineno}

\usepackage{graphicx}
\usepackage{subcaption}
\usepackage{multirow}
\usepackage[backend=biber, style=apa]{biblatex}
\usepackage{float}
\usepackage{etoolbox}
\usepackage[flushleft]{threeparttable}
\restylefloat{table}
\usepackage{setspace}
\title{Exploring Turn Signal Usage Patterns in Lane Changes: A Bayesian Hierarchical Modelling Analysis of Realistic Driving Data}

\author{ \href{https://orcid.org/0000-0003-1836-0664}{\includegraphics[scale=0.06]{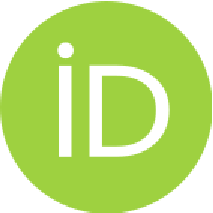}\hspace{1mm}Sarang Jokhio}\thanks{Corresponding author} \\
        Department of Human Factors, Institute of Psychology and Education \\
        Ulm University, 89089, Ulm, Germany\\
	\texttt{sarang.jokhio@uni-ulm.de}
	\And
	{Pierluigi Olleja} \\
	Division of Vehicle Safety, Department of Mechanics and Maritime Sciences \\
	Chalmers University of Technology, 	412 96, Gothenburg, Sweden 
        \And
	{Jonas Bärgman } \\
	Division of Vehicle Safety, Department of Mechanics and Maritime Sciences\\
	Chalmers University of Technology, 412 96, Gothenburg, Sweden
	  \And
        {Fei Yan} \\
	Department of Human Factors, Institute of Psychology and Education\\
        Ulm University, 89089, Ulm, Germany 
        \And
        {Martin Baumann} \\
		Department of Human Factors, Institute of Psychology and Education\\
        Ulm University, 89089, Ulm, Germany  
}

\renewcommand{\undertitle}{Preprint}
\renewcommand{\shorttitle}{Turn signal usage during lane change}

\begin{document}
\maketitle
\newpage
\begin{abstract}
Using turn signals to convey a driver's intention to change lanes provides a direct and unambiguous way of communicating with nearby drivers. Nonetheless, past research has indicated that drivers may not always use their turn signals prior to starting a lane change.  In this study, we analyze realistic driving data to investigate turn signal usage during lane changes on highways in and around Gothenburg, Sweden. We examine turn signal usage and identify factors that have an  influence on it by employing Bayesian hierarchical modelling (BHM). The results showed that a turn signal was used in approximately 60\% of cases before starting a lane change, while it was only used after the start of a lane change in 33\% of cases. In 7\% of cases, a turn signal was not used at all. Additionally, the BHM results reveal that various factors influence turn signal usage. The study concludes that understanding the factors that affect turn signal usage is crucial for improving traffic safety through policy-making and designing algorithms for autonomous vehicles for future mixed traffic.
\end{abstract}

\keywords{Lane change \and Turn signal usage \and Realistic driving data \and Bayesian hierarchical modeling \and Autonomous vehicles}

\section{Introduction}
The behaviour of drivers in traffic can be unpredictable, leading to complex and dynamic interactions that pose a challenge to the smooth functioning of autonomous vehicles (AVs). To be able to navigate and respond to human drivers safely and efficiently,  AVs should be programmed with the ability to anticipate and interpret human driver behaviour. Understanding and interpreting the various explicit and implicit signals that human drivers use in traffic is crucial for AVs to communicate effectively and interact with other road users in different situations. One of the interaction situations that can occur is during a lane change. The lane change is a frequently performed but complex manoeuvre that requires effective communication with other drivers to ensure safe completion. In a lane-changing scenario, a driver uses implicit and explicit forms of communication to convey the intentions to other road users \autocite{kauffmann_what_2018}. An implicit or indirect form of communication during a lane change occurs when the driver positions the vehicle in a certain way, such as moving closer to the lane boundary line, as an indication of their intention to change lanes. An explicit or direct form of communication during a lane change is using a turn signal to announce the driver's intentions to change lanes to surrounding drivers. 

A turn signal is a crucial safety feature that plays a vital role in effective communication between drivers, as it allows them to announce their intention to change lanes \autocite{ponziani_turn_2012}. As such, it is essential that turn signals are used properly and consistently, particularly before starting a lane change. Although the literature lacks a formal definition of ``proper turn signal use'' it is widely recognized that it involves activating the turn signal before starting a lane change. Using a turn signal before changing lanes allows drivers in the target lane to anticipate and prepare for the upcoming lane change by providing them with enough time to react and adjust their speed or position accordingly. Many countries require drivers to use turn signals before making a lane change, but there is a lack of clear regulations regarding the onset time (i.e., when it should be activated). For example, German road traffic law only requires drivers to signal the lane change intention "in good time" \autocite{StVO}, leaving room for interpretation. However, some states in the United States, such as California, have more specific regulations in road traffic law, requiring drivers to use a turn signal at least 5 seconds before a lane change \autocite{california_dmv_2020}. Proper turn signal usage is essential, as it could avoid many potential crashes \autocite{ponziani_turn_2012} and also increase cooperative behaviour (such as opening up gaps for lane changing vehicles) by drivers in the target lane \autocite{kauffmann_what_2018, stoll_situational_2020}. The use of proper turn signals is also imperative in the context of future mixed traffic scenarios, where AVs will coexist with conventional human-driven vehicles on shared roadways. This is because, as AVs rely heavily on strict rules, they may not understand the implicit forms of communication used by human drivers \autocite{farber2016communication}.

\textcite{tijerina_van_1997}, investigated turn signal usage of drivers while performing a lane change. The data for the study was collected on public roads in Ohio, USA, using an instrumented vehicle, with the driver being accompanied by an observer.  They reported that a turn signal was used 92\% of the time. However, the observers' presence in the passenger seat might have resulted in higher compliance than if no observers were present \autocite{lee_comprehensive_2004}. In a later study by \textcite{lee_comprehensive_2004}, 16 participants drove an instrumented vehicle on an interstate highway in Virginia, USA, without an observer. The study found that the turn signal was used only 44\% of the time. Turn signal usage for lane changes to the right was only 35\% compared to 48\% in the case of lane changes to the left lane. Unlike the above two studies, \textcite{ponziani_turn_2012} used an observer with two tally counters in a vehicle driving around Ohio, USA. The observer recorded the lawful use of the turn signal on one counter and unlawful or neglected on the second tally counter. A total of 2000 data points were gathered for lane changes. It was found that only 51.65\% of drivers complied with proper turn signal usage while making a lane change. \textcite{lin2019effect} used data collected in Michigan, USA, during the Safety Pilot field-operational-test sponsored by the National Highway Traffic Safety Administration \autocite{bezzina2015safety}. They reported the use of turn signals in about 70\%  of lane changes.

All of the above studies were from the USA. However, similar numbers were reported in the two studies conducted in China. For example, \textcite{dang2013analysis}, used an instrumented vehicle and studied 12 participants who drove approximately 200 Km (on each side) on a two-lane highway between the city of Beijing and Baoding. On the first leg of the journey (Beijing to Baoding), the drivers drove the vehicle with their usual driving behaviour. On the second leg of the journey (Baoding to Beijing), drivers were asked to make frequent lane changes and use turn signals before starting a lane change. The percentage of turn signal usage in the first leg was about 40\%. However, despite being instructed to use the turn signal before starting a lane change, the study found that its usage in the second leg of the drive was only about 65\%. The second study by \textcite{wang2019analysis}, used data collected during the Shanghai Naturalistic Driving Study. They analyzed lane changes made by the vehicle in the adjacent lane (both left and right) to the instrumented vehicle. After extracting the time of lane changes (called cut-in the original paper), they observed turn signal usage by watching forward roadway videos from the subject vehicles. The results showed that turn signal usage was less than 50\% on all types of roads, including freeways and expressways. These results indicate that turn  usage during lane change is relatively low, even though traffic law mandates it in many countries worldwide. 

As previously emphasized, the use of turn signals to indicate intent during a lane change is a vital aspect, yet it has received less attention than other characteristics, such as the duration of the lane change and acceptance of the gap.  Only a few studies on this topic have been conducted, and they predominantly originate from just two countries: the United States and China. To the authors' knowledge, there is a lack of research on turn signal behaviour, specifically in the context of drivers from European countries. The current literature also primarily relies on observations by a person in the vehicle or via videos recorded by instrumented vehicles.  Furthermore, there is a scarcity of research that differentiates between turn signals used before and after starting a lane change, except for the study by  \textcite{lin2019effect}.  Additionally, there is a gap in  extensive research exploring the various factors that could potentially impact the usage of turn signals during lane changes. Therefore, this study aims to examine the usage of turn signals among Swedish (European) drivers by analyzing data on their naturalistic behaviour during lane changes. Additionally, the paper provides an in-depth analysis of potential influencing factors on turn signal usage.

We used Bayesian hierarchical modelling to analyze the turn signal usage data. We used realistic driving data, which typically has a hierarchical or multilevel structure where observations are nested within participants. In recent years, Bayesian approaches have been increasingly applied in transportation and other fields such as social and behavioural sciences \parencite{russo2020statistics, depaoli2017introduction, daziano2013computational}. In the transportation domain, the Bayesian approach has been used to study road safety \autocite{huang2010multilevel}, such as crash estimation, prediction of road accidents, and road network safety evaluation \autocite{xie2013corridor, zheng2019bayesian, yu2014using,deublein2013prediction}. In addition, Bayesian techniques have been employed to develop reference models \autocite{morando2019bayesian} and to produce synthetic data that mimics driver behaviour \autocite{schindler2021making}.

The structure of this paper is as follows. Section  In Section \ref{data}, we thoroughly examine the data utilized in this study. In section \ref{bhm}, we discuss Bayesian hierarchical modelling, followed by a presentation of our findings in Section \ref{results}. Lastly, the implications of our results, potential limitations, and recommendations for future research are discussed in Section \ref{discussion}.
\section{Data Description}\label{data}
The data used in this paper were collected during the L3Pilot Project \autocite{penttinen2019experimental}. A dataset collected by Volvo Car Corporation for this project includes vehicles equipped with an automated driving function being driven on the highway surrounding the city of Gothenburg, Sweden. This dataset includes both trips where the driving task was purely manual (baseline) and trips where the automated driving function was activated (treatment). Drivers in the baseline dataset were primarily non-professional drivers. The professional drivers were involved in part of the baseline and in the whole treatment dataset. In the context of L3Pilot, the term "professional" referred to drivers with a qualification in operating prototype vehicles for testing specific features \autocite{penttinen2019experimental,BundesagenturArbeit2018}. Professional drivers were also compensated for driving. Non-professional drivers, on the other hand, are individuals who are recruited within the company \autocite{penttinen2019experimental}. In our study, only baseline trips with both non-professional and professional drivers were considered. The cars were equipped with cameras facing outside towards traffic (Front and Rear) and inside towards the driver, radars, accelerometers and angular rate sensors, and a global positioning system (GPS). In addition, data from the Control Area Network (CAN) bus in the vehicle was also collected, including information such as turn indicator usage and ego vehicle speed. Furthermore, the longitudinal and lateral positions, the speed of surrounding vehicles, and traffic density were also collected. In this study, we used data from the ego vehicle described below to determine the start and completion of a lane change manoeuvre.

Determining the start and completion of a lane change from time-series data can be challenging since the measured values are highly sensitive to the definitions selected by authors \autocite{wang2014investigation}. As a result, researchers often use various methods to identify lane change start and completion in their data. What can and cannot be done primarily depends on which data is available \autocite{xi2013review}. The thresholds of lateral distance to a lane marking and the lateral velocity of the vehicle are common variables used to determine a lane change's start and completion point. This approach has been used on data collected from vehicle-mounted cameras \autocite{olsen2002analysis} and data from high-rise building-mounted cameras \autocite{toledo_modeling_2007}. However, neither study provided quantitative thresholds for lateral position or velocity. Thresholds provide objective, quantifiable criteria to determine lane change start and endpoints. For example, \textcite{wang2014investigation}, used a -0.2 m/s and \textcite{mullakkal-babu_empirics_2020}, used a 0.33 m/s lateral velocity threshold to determine the start  and end of a lane change.

In this study, we first filtered lane change scenarios using L3Pilot's scenario definition, which only considered the distance to lane markings \autocite{penttinen2019experimental}. However, we used thresholds of lateral velocity and distance to lane markings as criteria to determine lane changes' start and completion points. To explore the appropriate thresholds, we first analyzed the driver's lateral velocity and position within the lane in a free-driving (the subject vehicle following its path without any influence of other vehicles) scenario. We found that in 90\% of free driving time, the lateral speed did not exceed 0.17 m/s both driving towards the left and the right. Moreover, the lateral position within the lane was found to be centred in the middle of the lane and to be within 0.42 m to the right and 0.36 m to the left 90\% of the time. 

Assuming that the lateral velocity at the start and completion of the lane change is zero \autocite{li_comprehensive_2021}, or using a smaller threshold, may result in underestimating the start and completion of lane changes. On the other hand, using a larger threshold could result in identifying the start or completion of lane change after the vehicle has already crossed the lane boundary. Therefore, we tested different thresholds of lateral velocity to determine the start and completion of a lane change. We selected five thresholds from 0.10 m/s to 0.3 m/s (with an increase of 0.05 m/s) in the lane change scenarios. We also verified random cases for each chosen threshold by watching a front-facing video of the lane-changing vehicle to see if the lane change occurred at a particular threshold. We found that vehicles already crossed the lane boundary while using a threshold of 0.20 m/s to 0.3 m/s. Finally, we decided to use a threshold of 0.15 m/s to determine the start of a lane change. We selected this threshold to strike a balance between instances where a vehicle may have already crossed lanes and instances where a vehicle is fluctuating within its current lane. To cover the situations where a lane change might happen slowly without reaching the lateral velocity criteria, we also included cases with a lateral motion that got closer than 5 cm from the lane marking at a start of a lane change. It is important to note that the study's focus was exclusively on the start of lane changes. Therefore, all variables used in the model were extracted at the start of the lane change, as defined by the 0.15m/s lateral speed, complemented with the 5cm-from-lane-marking  criterion.

\subsection{Descriptive Data}

A total of 1791 lane change cases for 103 drivers were extracted from the dataset, having met either of the previously described thresholds. The driver's age ranged from 23 to 64 years old, with an average of 40.2 years old. The drivers consisted of 71 males and 32 females. Among these drivers, 69\% had more than ten years of driving experience, while the remaining drivers had shorter or non-specified driving experience. 

\begin{figure}[h]
    \centering
    \includegraphics[width=130mm]{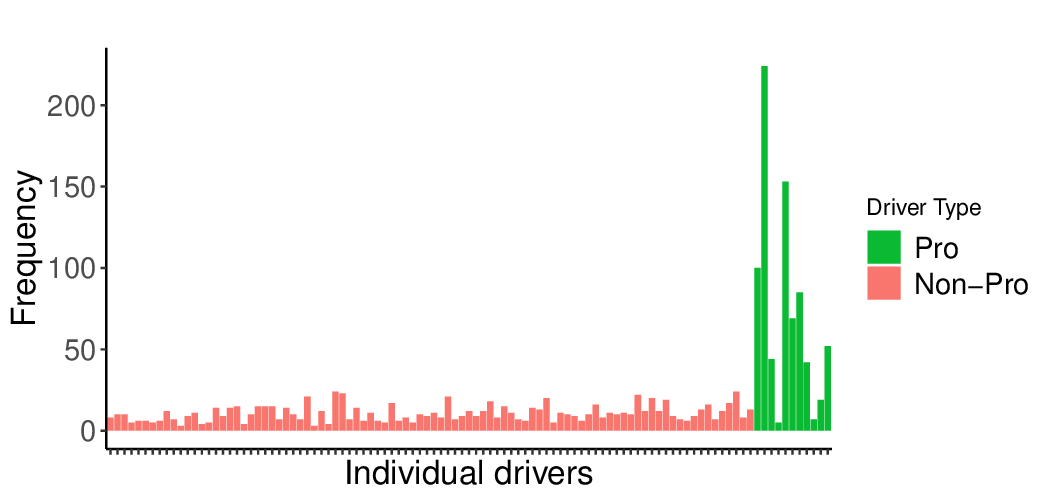}
    \caption{The frequency of LC cases per driver}
    \label{fig1}
\end{figure}

As shown in Figure \ref{fig1}, there is an imbalance in the number of lane changes per driver. Of the 103 drivers in the dataset, only eleven (about 11\%) were professional drivers. Despite this, they accounted for nearly 40\% of the lane changes. This was due to the fact that professional drivers drove more and thus generated more driving data compared to non-professional drivers who typically drove for only one hour. As mentioned earlier, the exclusion of trips where the automated driving function was activated has led to an imbalance among the trips of professional drivers. Furthermore, although the non-professional drivers drove for a similar duration, they exhibited variation in their driving patterns. This variation can be attributed to the fewer situations where a lane change was required.

\begin{figure}[h]
    \centering
    \includegraphics[width=120mm]{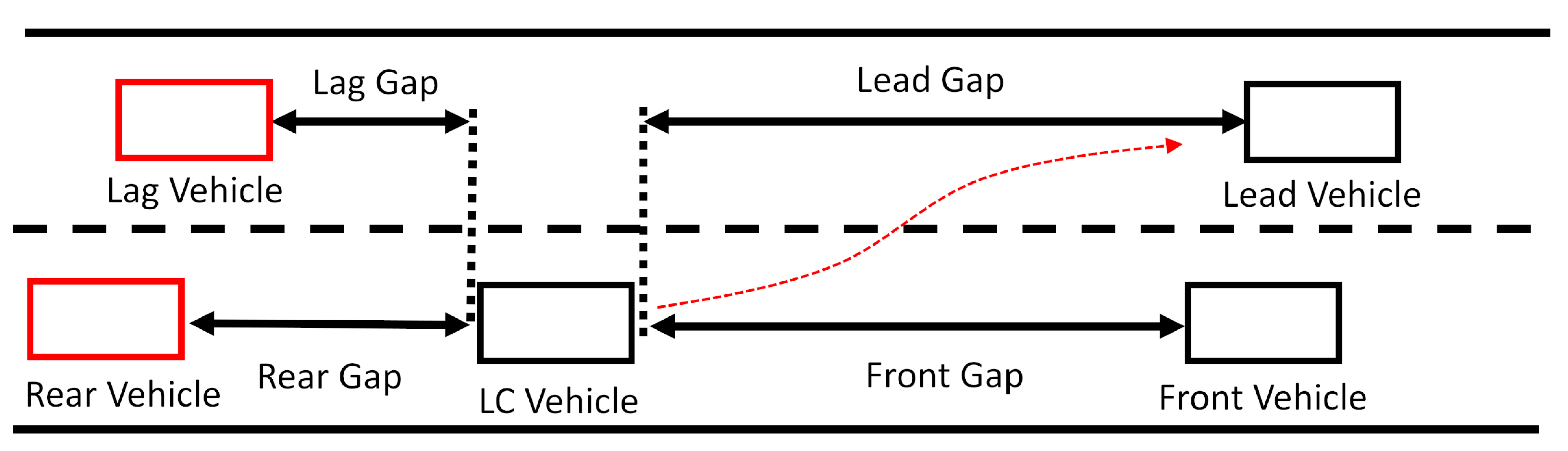}
    \caption{A typical LC Scenario}
    \label{fig2}
\end{figure}

Figure \ref{fig2} illustrates a typical lane change scenario, which depicts the different vehicles involved. The vehicles shown in red (rear and lag vehicles) potentially impact the turn signal usage behavior of the lane-changing vehicles (shown in blue). This is because the turn signal is primarily used to communicate to the rear and lag vehicles that a lane change is imminent.

Overall turn signal usage was 92.8\%, which is higher than reported in previous literature. However, further examination revealed a discrepancy in turn signal usage, specifically in relation to lane change start. In 59.9\% of cases, a turn signal was used before starting a lane change, while it was only used after the start of a lane change in 32.9\% of instances. Furthermore, in 7.2\% of cases, no turn signal was used at all. 

The following variables were taken into consideration for each lane change case: speed (km/h) of the ego vehicle, direction of lane change (right or left), presence of a rear vehicle (yes or no), rear gap (seconds), presence of a lag vehicle (yes or no), lag gap (seconds), and traffic density (vehicles/km/lane) and driver type (pro and no pro). The variables were selected because of their potential effect on turn signal usage. For instance, it was demonstrated in \cite{lee_comprehensive_2004} that turn signal usage varied depending on the direction of the lane change. Research has also shown that surrounding vehicles and their respective gaps impact the overall lane change process \autocite{moridpour2010effect}, although not directly on turn signal usage.

\section{Bayesian Hierarchical Model}\label{bhm}
The realistic driving data, such as the one we used, are often  large, noisy, sparse, and unbalanced \autocite{morando2019bayesian}. Furthermore, it typically has a hierarchical or multilevel structure, where observations are nested within participants. To account for the variation,  hierarchical models such as mixed effect multinomial Logit models are typically used in the traditional approach \autocite{bhat2004mixed}. Nonetheless, hierarchical models based on the traditional approach have some limitations  that can be addressed by employing a Bayesian approach. Bayesian data analysis provides a more intuitive interpretation of results using probabilities as opposed to p-values in traditional approach. The bayesian approach can handle more complex and non-linear relationships, as well as non-normally distributed data \autocite{kruschke2018bayesian}.

At its fundamental level, the Bayesian approach uses Bayes' theorem to update prior beliefs or knowledge with observed data to make inferences about unknown parameters \autocite{van2021bayesian}. Bayes' theorem is the fundamental rule that combines prior probability and likelihood to compute the posterior probability. The prior probability represents the initial belief about the parameters or hypotheses before observing any data. In contrast, the posterior probability is the updated belief about the parameters after observing the data. For an in-depth introduction to Bayesian data analysis, readers are referred to \autocite{gelman2013bayesian}.

Bayesian hierarchical (or multilevel) models (BHM) are a class of statistical models that allow for the modelling of complex relationships within hierarchical or nested data structures \autocite{gelman2006data}.  In our case, the data for lane change measurements are obtained from various drivers. This data has a multilevel structure, which makes BHM an advantageous statistical modelling approach to capture the intricate relationships within the nested data structure. Although not unique to BHM, a key feature of BHM is that they allow for the estimation of both fixed and random effects, enabling the decomposition of variance in the response variable into components attributable to different levels of the hierarchy. In this study, the probability of observing each level of the outcome variable (Signal) is modelled using equation \ref{eq1} and \ref{eq2}

\begin{equation}\label{eq1}
\mathrm{logit}\left(\frac{P(\mathrm{Signal} = \mathrm{after})}{P(\mathrm{Signal} = \mathrm{before})}\right) = \ \alpha_{\mathrm{after}} + \beta_{1,\mathrm{after}} \times X + b_{0j,\mathrm{after}}
\end{equation}

\begin{equation}\label{eq2}
\mathrm{logit}\left(\frac{P(\mathrm{Signal} = \mathrm{no})}{P(\mathrm{Signal} = \mathrm{before})}\right) = \ \alpha_{\mathrm{no}} + \beta_{1,\mathrm{no}} \times X + b_{0j,\mathrm{no}}   
\end{equation}


Here, X represents the vector of predictor variables, $\alpha_{\mathrm{after}}$ and $\alpha_{\mathrm{no}}$ are the intercepts for the "after" and "no" categories, respectively.
$\beta_{1,\mathrm{after}}$ and $\beta_{1,\mathrm{no}}$ represent the fixed effect coefficients for the "after" and "no" categories, respectively, corresponding to the vector X of predictor variables.
In these equations, $b_{0j,\mathrm{after}}$ and $b_{0j,\mathrm{no}}$ represent the random intercepts for each driver.

The categorical response variable Signal has three levels: before, after, and no. The level "before" is chosen as the baseline. The model will estimate the person-specific effects of the predictors and the random intercept for each driver. In summary, the model includes fixed effects for speed, direction, rear vehicle, rear gap, lag vehicle, lag gap, traffic density, driver type, and random intercepts for  Driver ID. The random intercept allows us to account for any variation across different drivers. The hierarchy of the model is explained below.

\begin{itemize}
    \item Random effects: At this level, the focus is on the individual drivers represented by the variable Driver ID. The random intercept ($b_{0j}$) accounts for the differences in the baseline of the response variable (Signal) across different drivers. This allows the model to capture variations between drivers, which can help understand individual drivers' unique characteristics and behavior.

    \item Fixed effects: At this level of the model, the response variable (Turn signal usage) is modelled as a function of the fixed effects, which include the predictors' (speed, direction, rear veh, rear gap, lag veh, lag gap, traffic density, and driver type). These fixed effects represent the average effects of each predictor on the response variable across all drivers. By including these fixed effects, the model captures the general relationship between the predictors and the response variable, allowing for an understanding of the overall trends in the data.
\end{itemize}

Integrating prior knowledge into the estimation process is a key advantage of Bayesian approach \autocite{depaoli2020importance}. A prior is a probability distribution that expresses the degree of belief or uncertainty about the value of a parameter of interest before any data is observed. The prior distribution is used as a starting point to compute the posterior distribution, which represents the updated degree of belief or uncertainty about the parameter after observing the data. Nevertheless, this feature is also a source of common criticism, as there is no unique way of choosing a prior distribution, and prior specification can substantially influence the model outcomes \autocite{robert2007bayesian,depaoli2020importance}. Without prior knowledge, often weakly informative or non-informative priors are assumed \autocite{gelman2013bayesian}. Weakly informative priors are sometimes used when some prior information about the parameters (e.g., the direction of effect) is available. Still, it is not strong enough to justify a prior solid distribution. Conversely, non-informative priors are often used when no prior information is available or the analyst wants the data to determine the posterior distribution completely. 

The BHM approach allows the use of priors while reducing the model's sensitivity to the choice of the prior distribution \autocite{robert2007bayesian}. In BHM, the model parameters are organized into multiple levels or layers, each with its own prior distribution. In BHM, we  incorporate additional information or structure into the model, which can help reduce the influence of the prior distribution on the resulting inference. 

In our study, we used non-informative priors with normal distribution $\mathcal{N}(0, 1\times10^4)$ for the fixed effects. This decision was made because there is no credible information available regarding the size or direction of the effects of different factors on turn signal usage. By using noninformative priors, we allow the data to influence the estimates most while still incorporating the prior information in the Bayesian framework. The priors for the random effects, such as for the standard deviations of the ``after'' and ``no'' parameters within each Driver ID group, are specified as half-normal $\text{Half-normal}(0, 1\times10^4)$ distributions.  Half-normal distributions are suitable for random effects because they are non-negative and non-informative \autocite{Gelman2006}. They have minimal influence on the results while providing some regularization to improve convergence and sampling efficiency.

The model is implemented using brms \autocite{burkner2017brms} package of R programming language \autocite{RCoreTeam2022}, which  provides an interface to fit Bayesian models using Stan. Stan is an open-source state-of-the-art platform for statistical modelling and high-performance statistical computation \autocite{carpenter2017stan}. The model uses Markov Chain Monte Carlo (MCMC) sampling with four separate chains, where each chain comprises 10,000 iterations, including an initial 5,000 burn-in iterations. During the burn-in phase, the sampler adapts and fine-tunes its parameters.

\section{Results}\label{results}
\begin{figure*}[h!]
\centering
\begin{subfigure}[t]{0.22\textwidth}
\includegraphics[width=\textwidth]{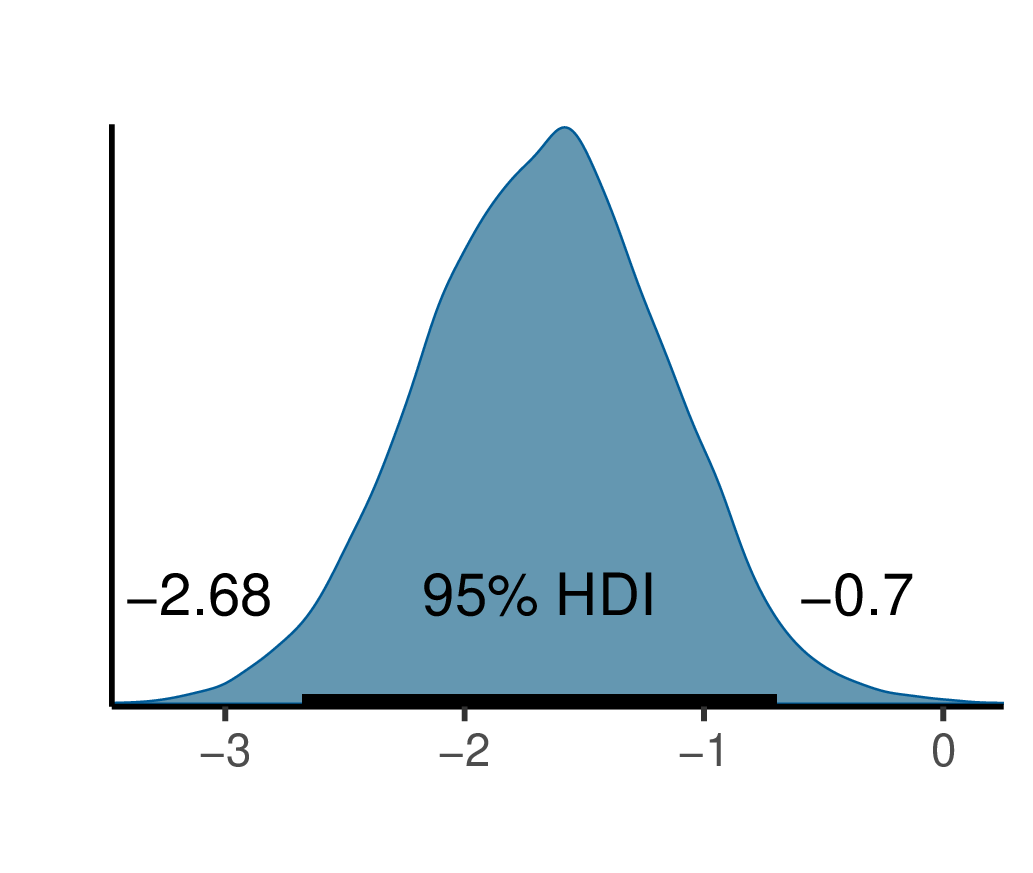}
\caption{Intercept (after)}
\label{fig5:sub1} 
\end{subfigure}
~
\begin{subfigure}[t]{0.22\textwidth}
\includegraphics[width=\textwidth]{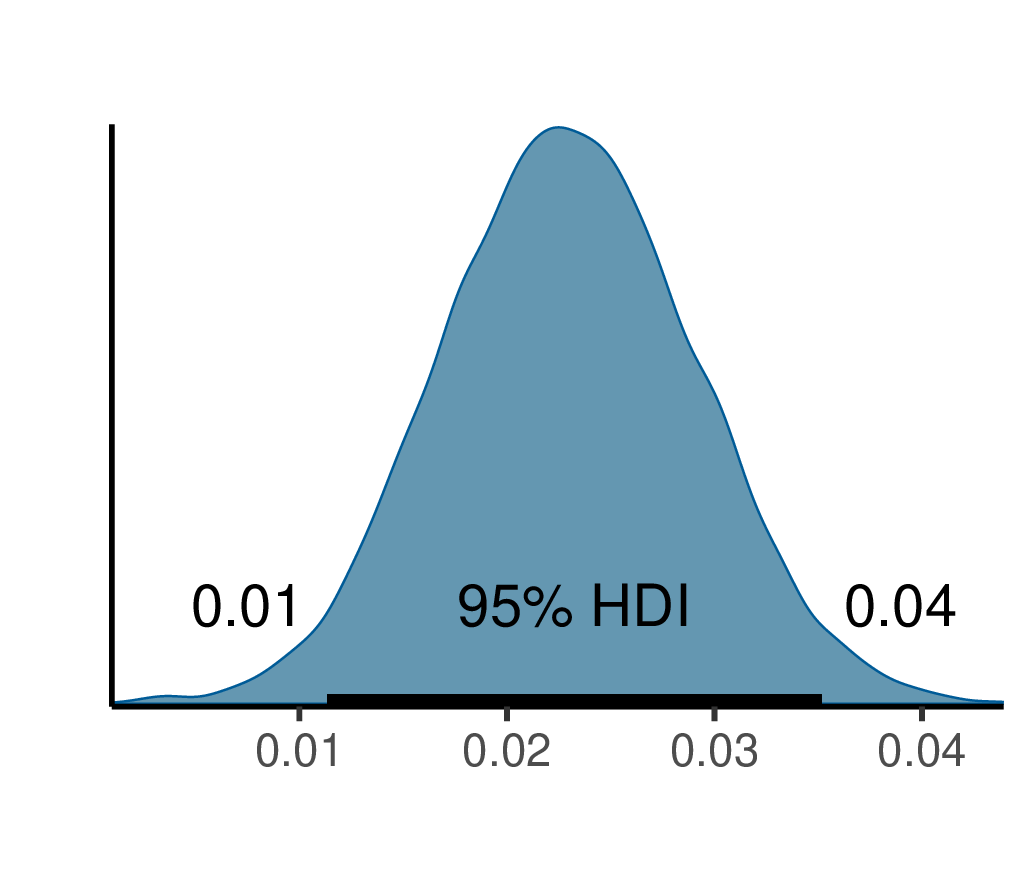}
\caption{Speed (after) }
\label{fig5:sub2}
\end{subfigure}%
~
\begin{subfigure}[t]{0.22\textwidth}
\includegraphics[width=\textwidth]{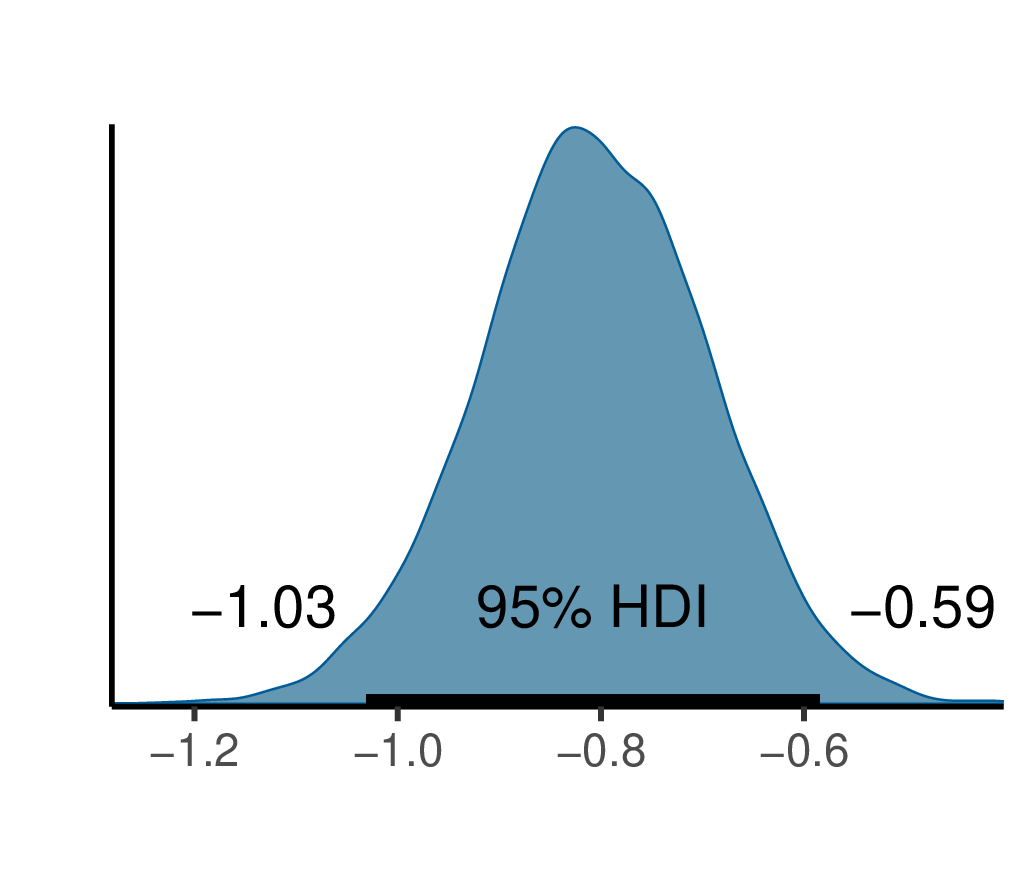}
\caption{Direction (after)}
\label{fig5:sub3}
\end{subfigure}
~
\begin{subfigure}[t]{0.22\textwidth}
\includegraphics[width=\textwidth]{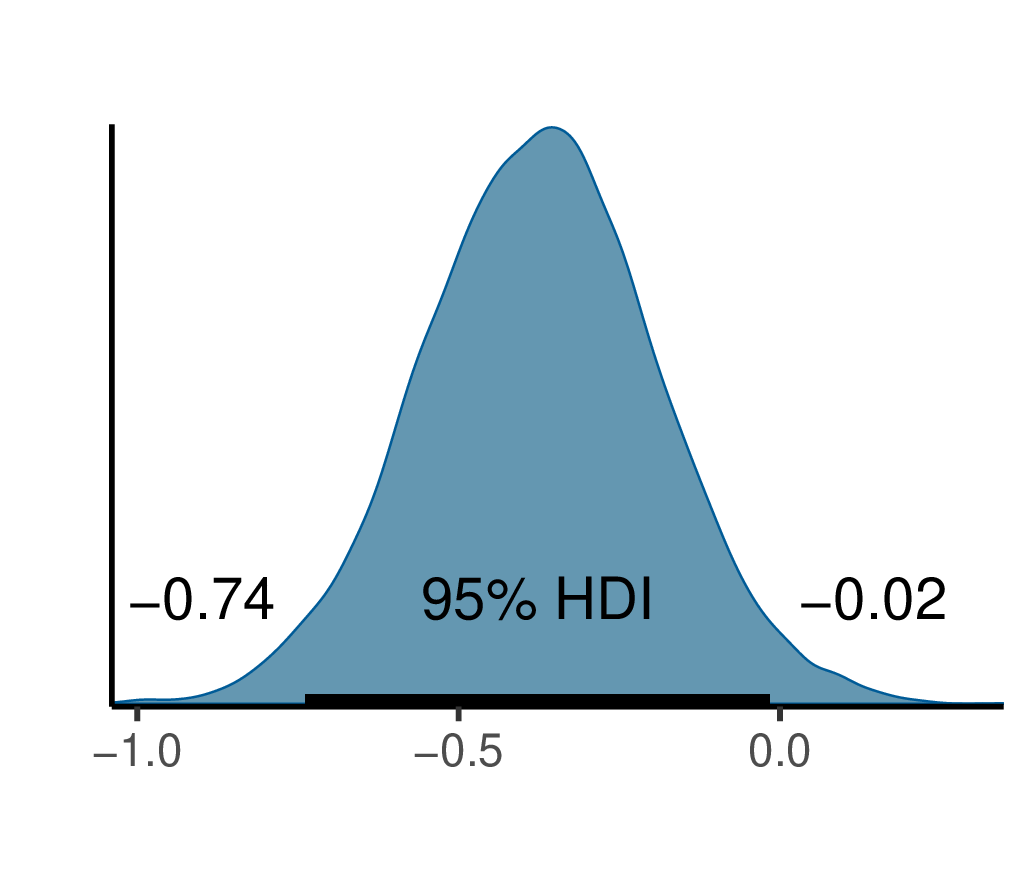}
\caption{Rear vehicle (after)}
\label{fig5:sub4}
\end{subfigure}
~
\begin{subfigure}[t]{0.22\textwidth}
\includegraphics[width=\textwidth]{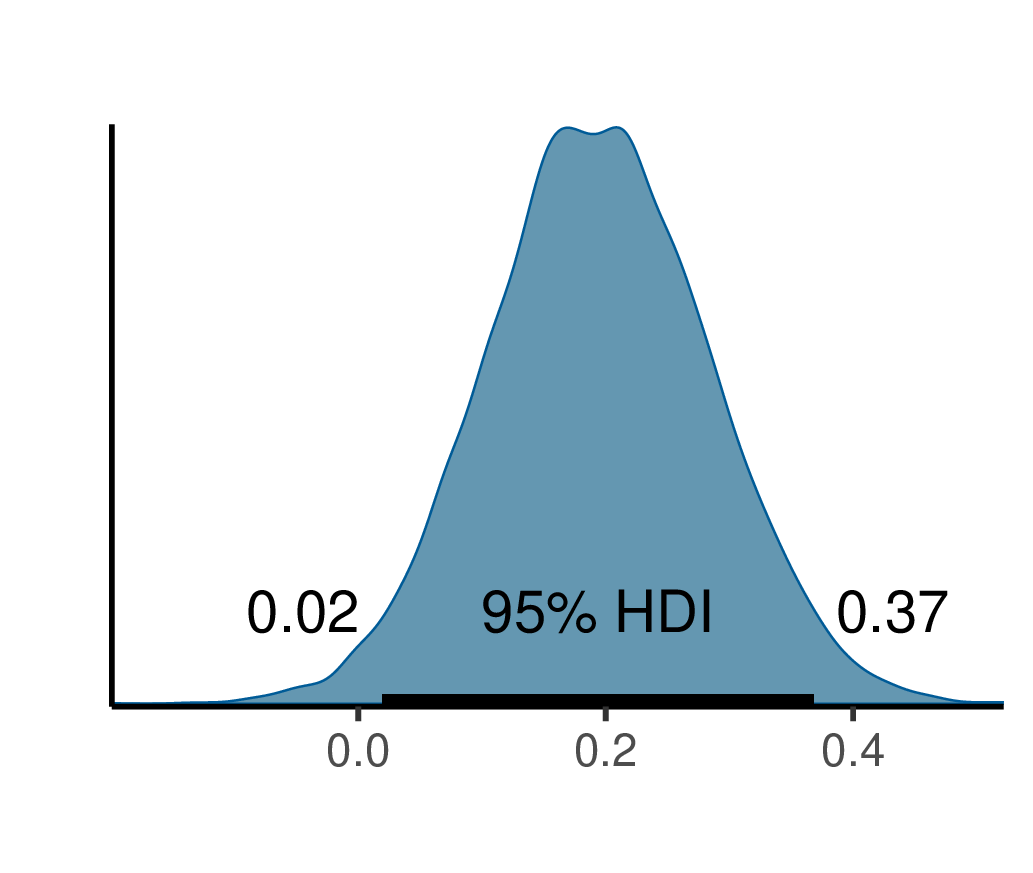}
\caption{Rear gap (after)}
\label{fig5:sub5}
\end{subfigure}
~
\begin{subfigure}[t]{0.22\textwidth}
\includegraphics[width=\textwidth]{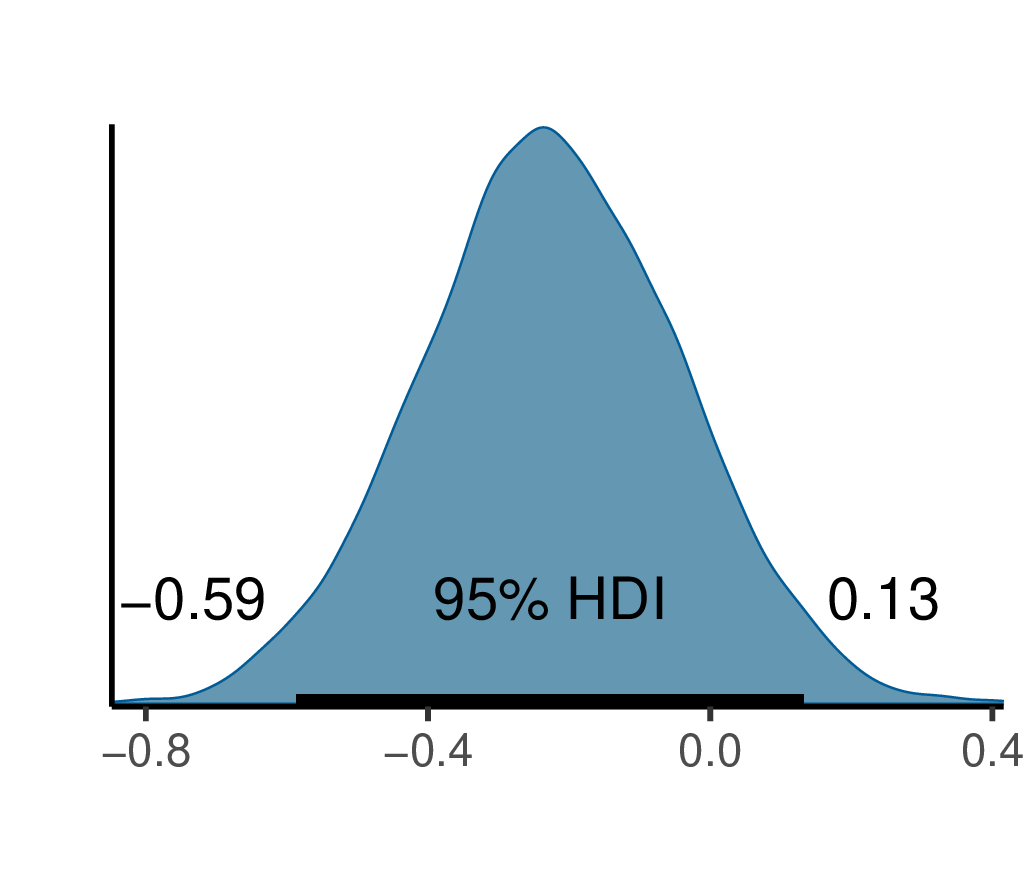}
\caption{Lag vehicle (after)}
\label{fig5:sub6}
\end{subfigure}
~
\begin{subfigure}[t]{0.22\textwidth}
\includegraphics[width=\textwidth]{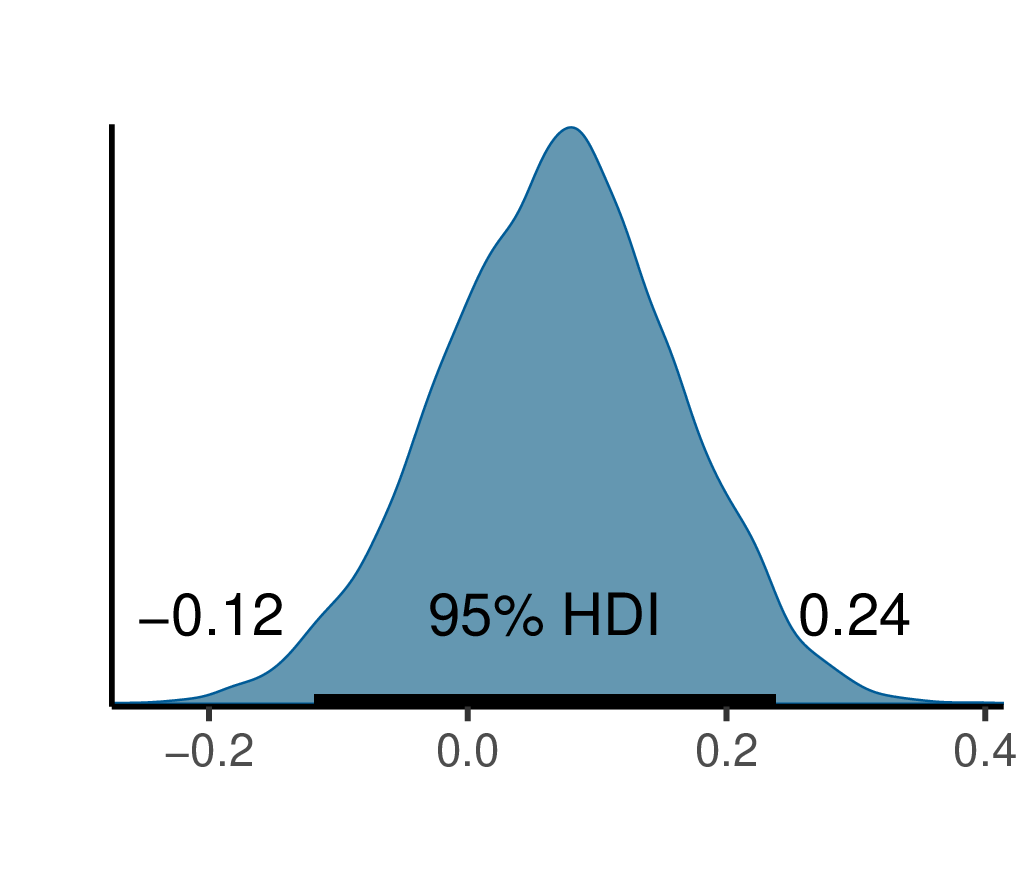}
\caption{Lag gap (after)}
\label{fig5:sub7}
\end{subfigure}
~
\begin{subfigure}[t]{0.22\textwidth}
\includegraphics[width=\textwidth]{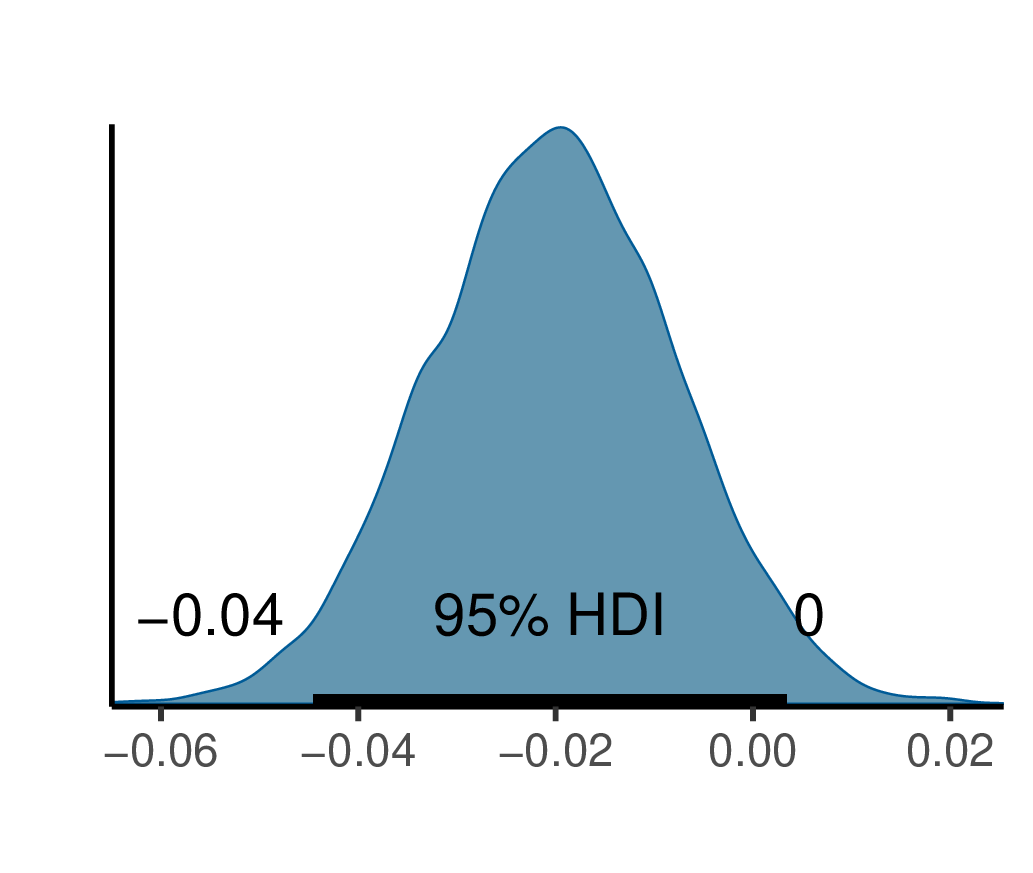}
\caption{Traffic density (after)}
\label{fig5:sub8}
\end{subfigure}
~
\begin{subfigure}[t]{0.22\textwidth}
\includegraphics[width=\textwidth]{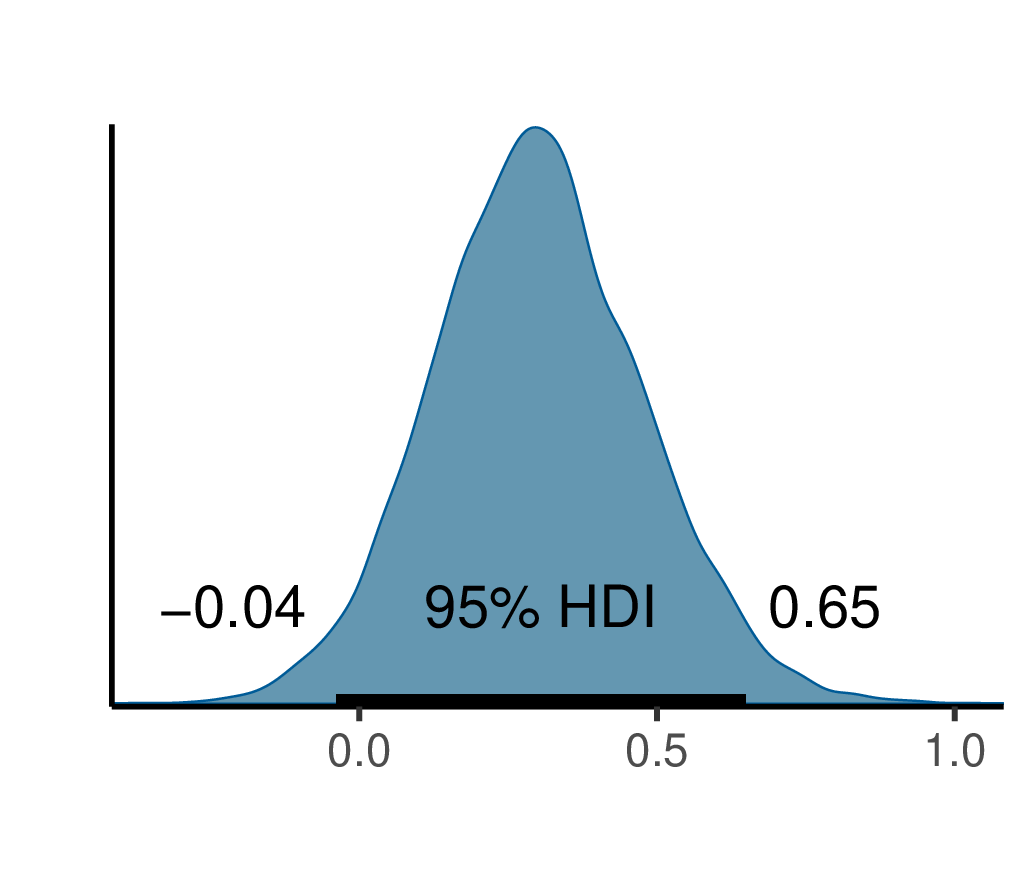}
\caption{Traffic density (after)}
\label{fig5:sub9}
\end{subfigure}
~
\begin{subfigure}[t]{0.22\textwidth}
\includegraphics[width=\textwidth]{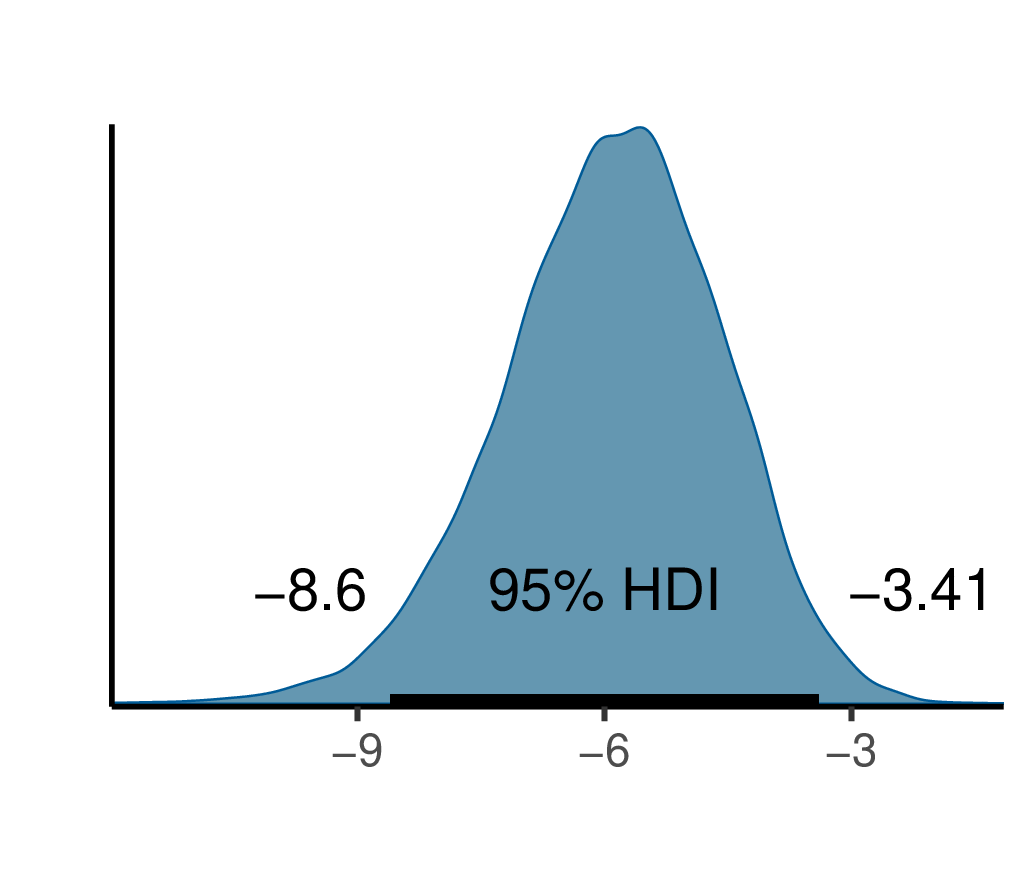}
\caption{Intercept (no)}
\label{fig5:sub10}
\end{subfigure}
~
\begin{subfigure}[t]{0.22\textwidth}
\includegraphics[width=\textwidth]{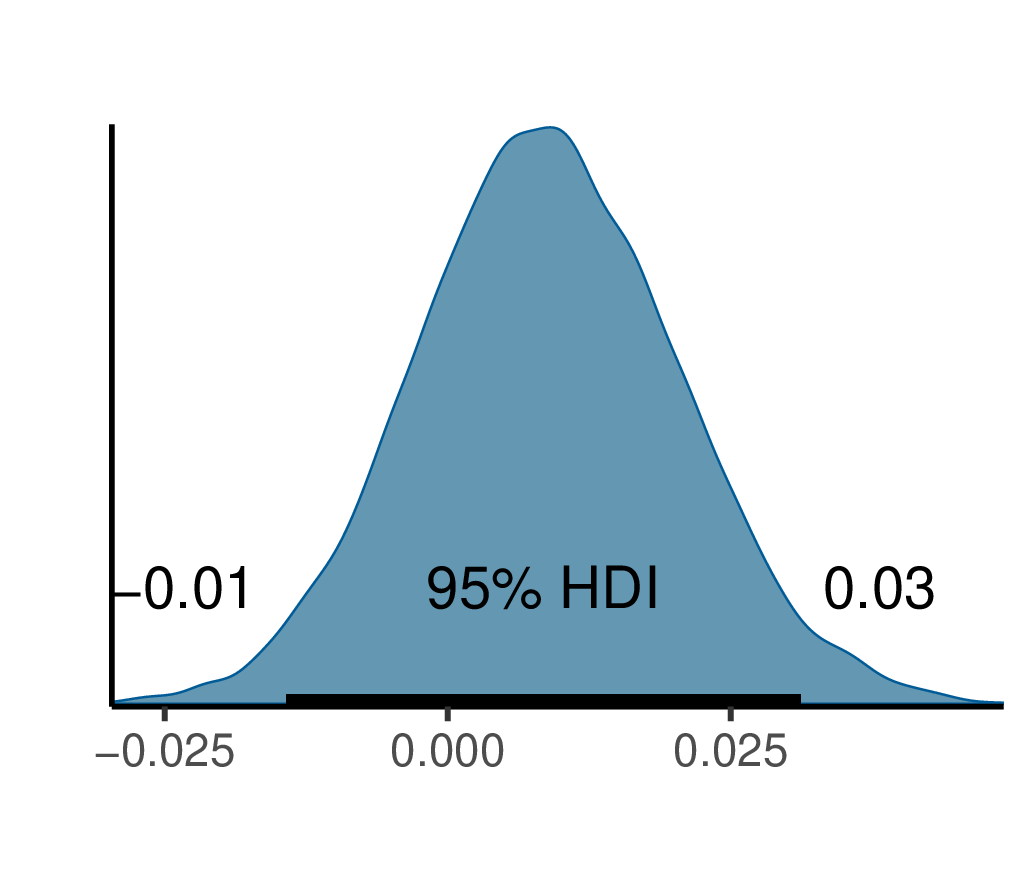}
\caption{Speed (no)}
\label{fig5:sub11}
\end{subfigure}
~
\begin{subfigure}[t]{0.22\textwidth}
\includegraphics[width=\textwidth]{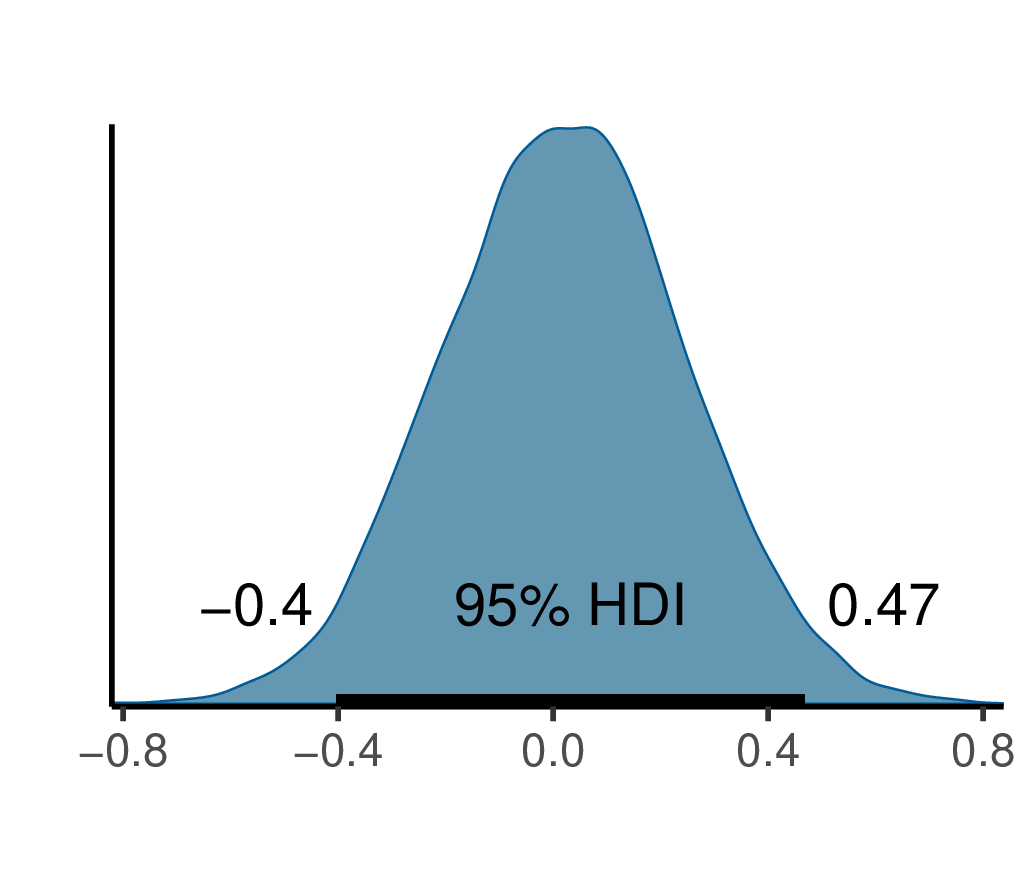}
\caption{Direction (no)}
\label{fig5:sub12}
\end{subfigure}
~

\begin{subfigure}[t]{0.22\textwidth}
\includegraphics[width=\textwidth]{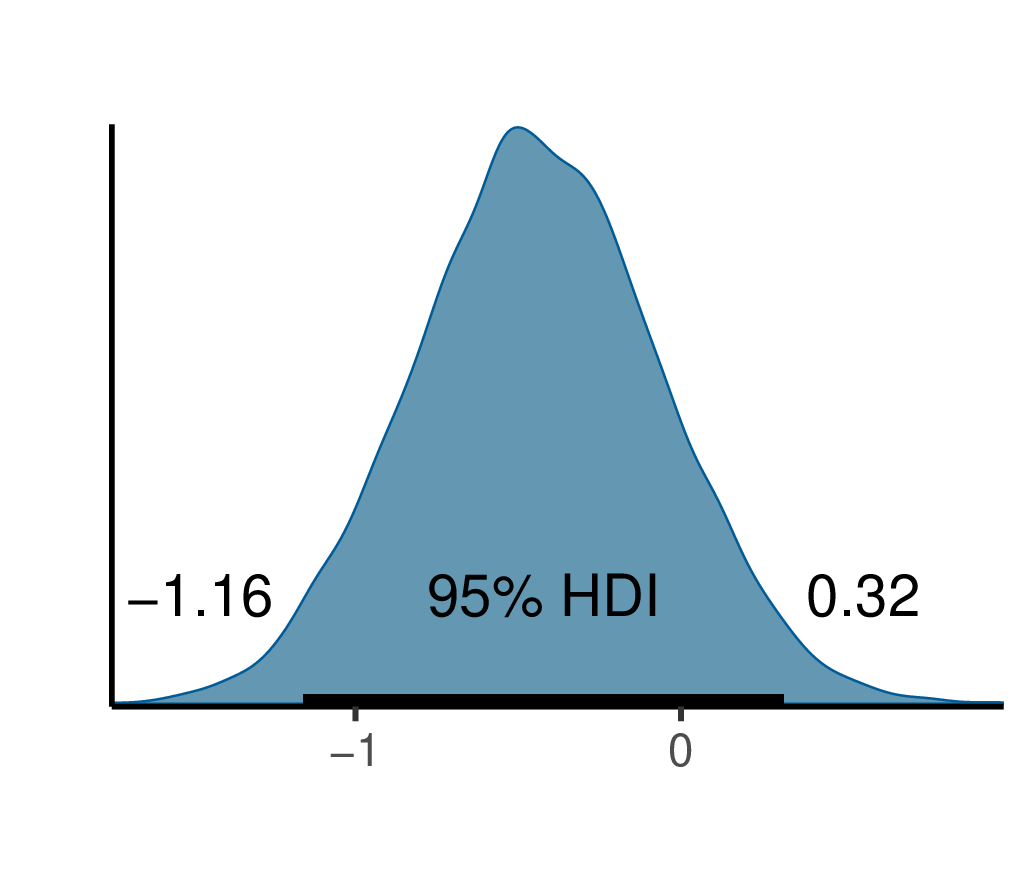}
\caption{Rear vehicle (no)}
\label{fig5:sub13}
\end{subfigure}
~
\begin{subfigure}[t]{0.22\textwidth}
\includegraphics[width=\textwidth]{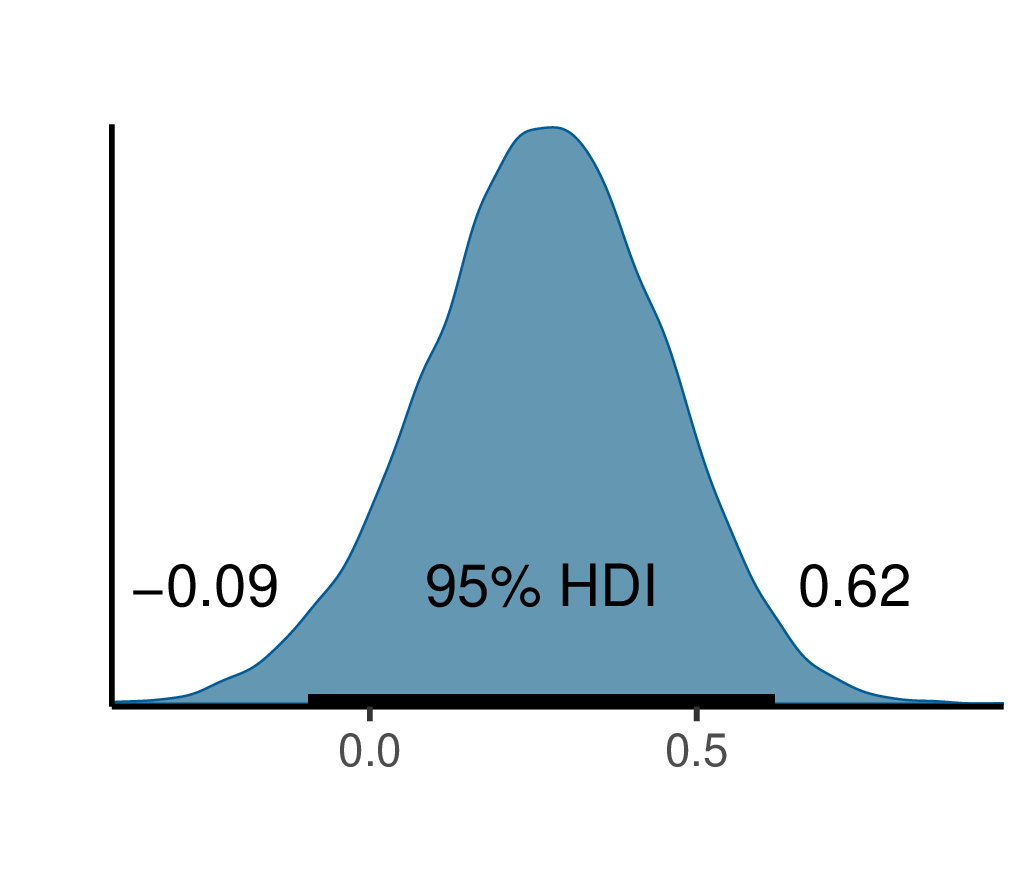}
\caption{Rear gap (no)}
\label{fig5:sub14}
\end{subfigure}
~
\begin{subfigure}[t]{0.22\textwidth}
\includegraphics[width=\textwidth]{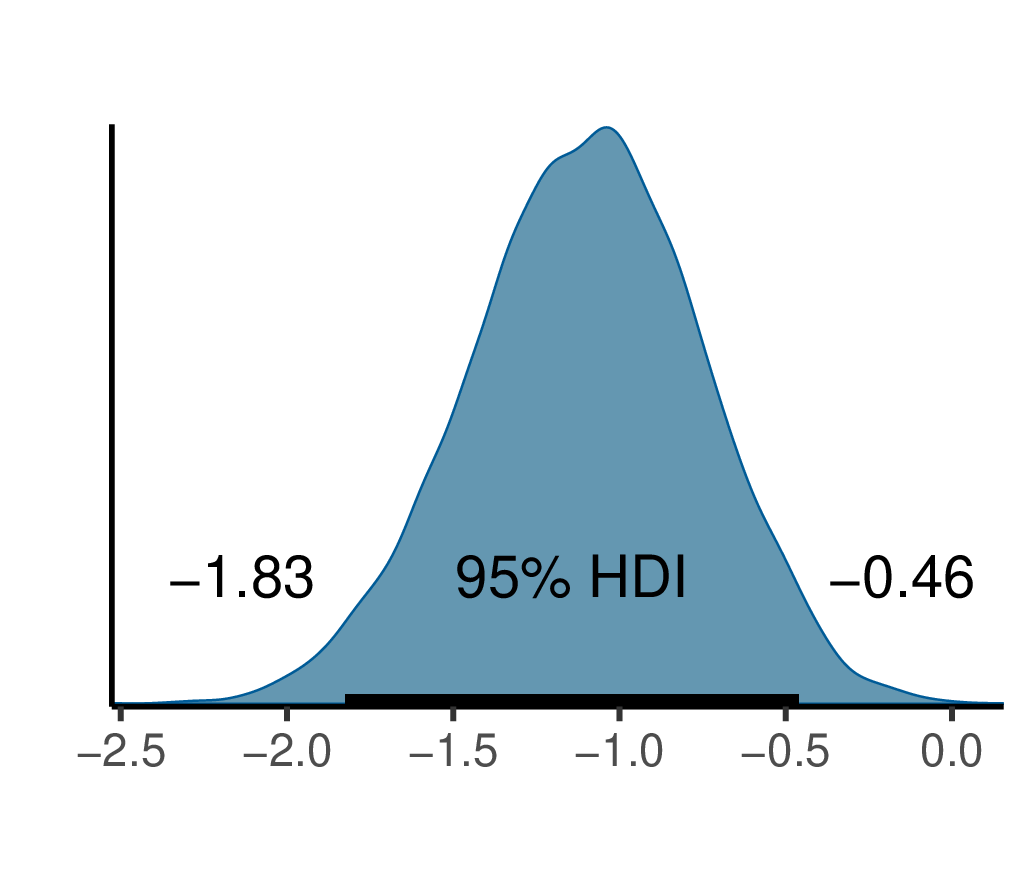}
\caption{Lag vehicle (no)}
\label{fig5:sub15}
\end{subfigure}
~
\begin{subfigure}[t]{0.22\textwidth}
\includegraphics[width=\textwidth]{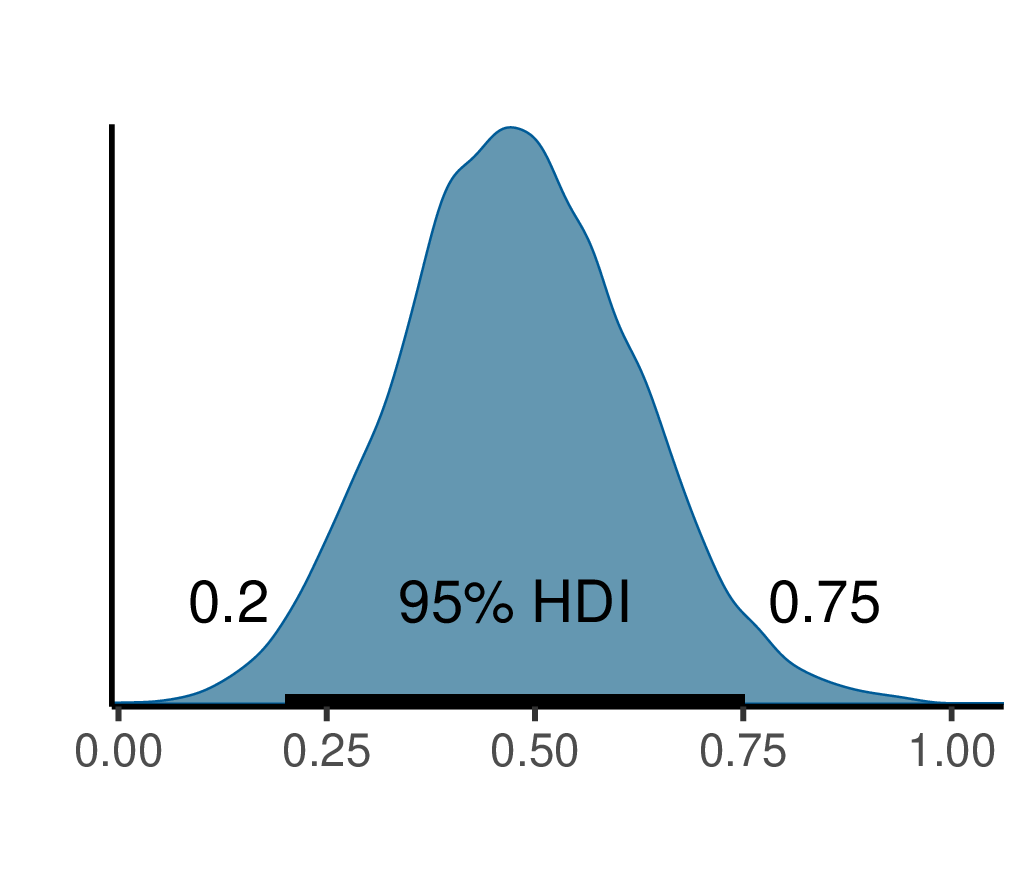}
\caption{Rear gap (no)}
\label{fig5:sub16}
\end{subfigure}
~
\begin{subfigure}[t]{0.22\textwidth}
\includegraphics[width=\textwidth]{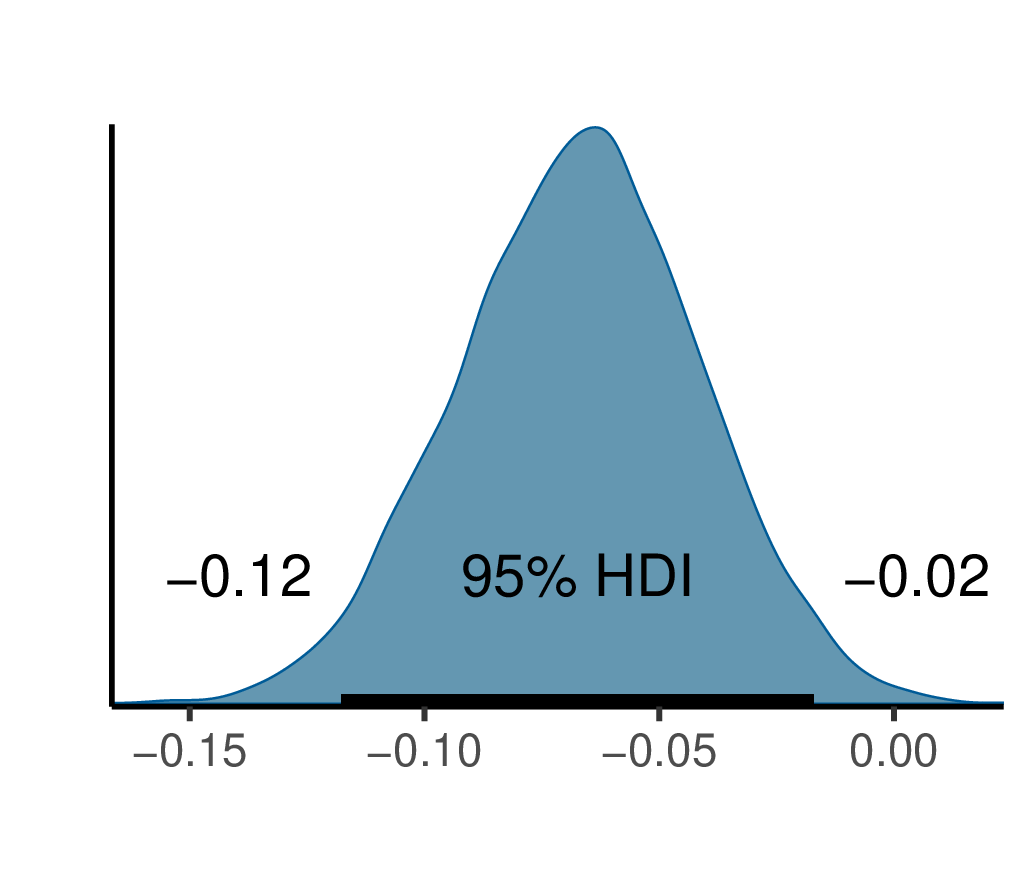}
\caption{Traffic density (no)}
\label{fig5:sub17}
\end{subfigure}
~
\begin{subfigure}[t]{0.22\textwidth}
\includegraphics[width=\textwidth]{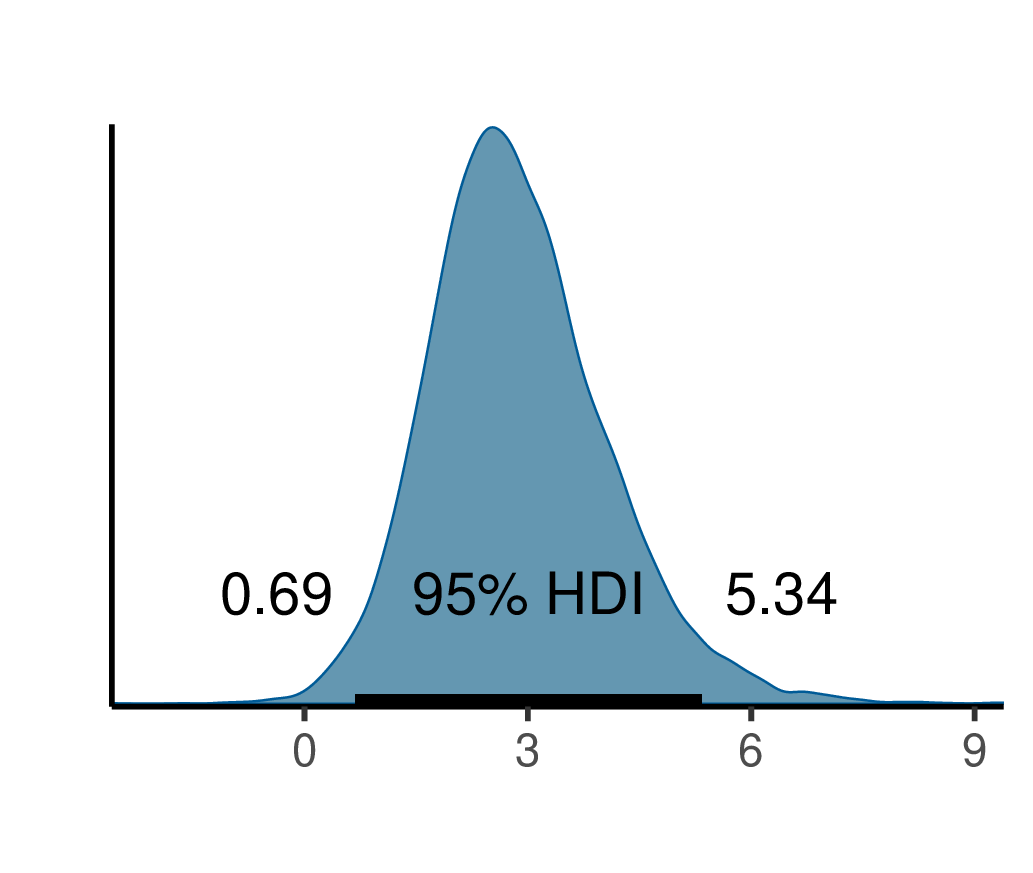}
\caption{Driver type (no) }
\label{fig5:sub18}
\end{subfigure}
~
\begin{subfigure}[t]{0.23\textwidth}
\includegraphics[width=\textwidth]{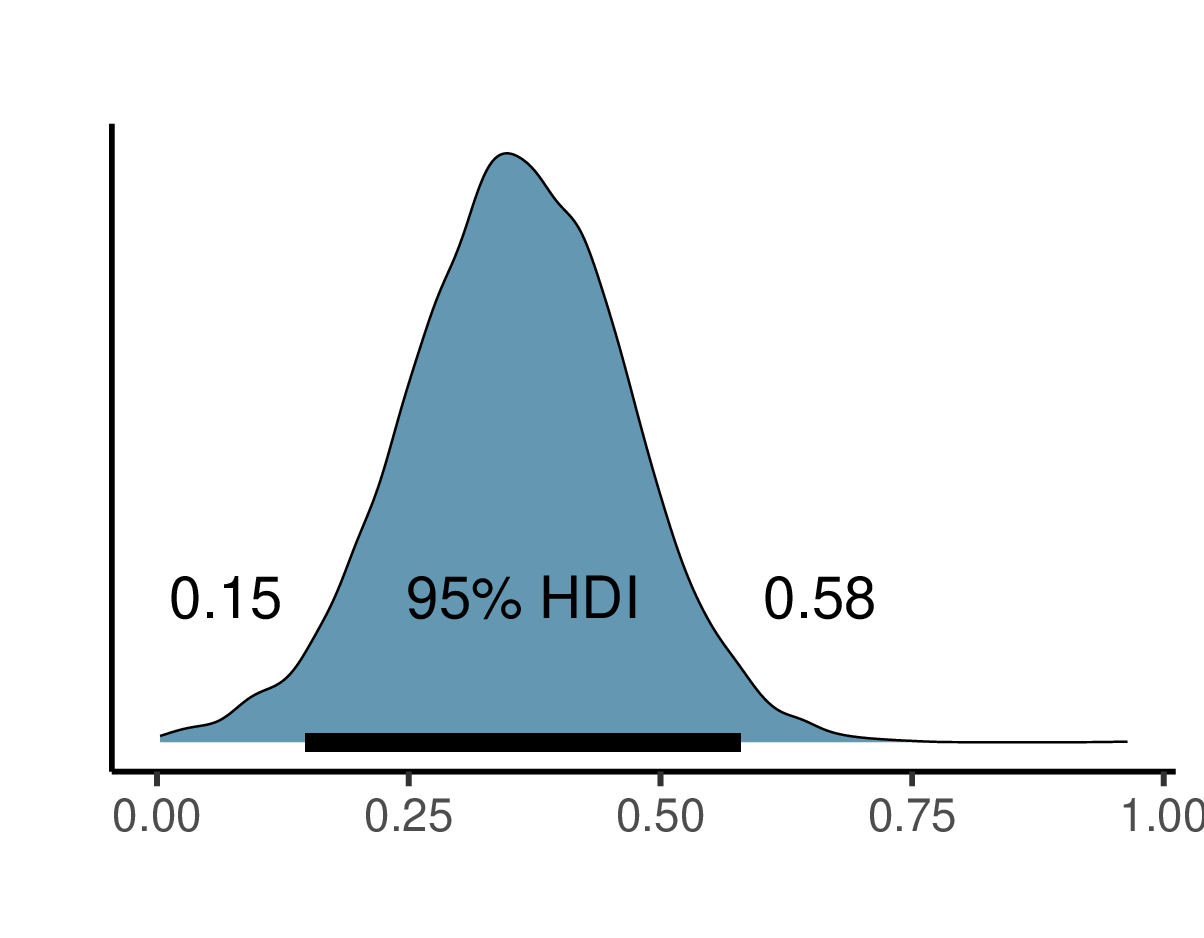}
\caption{Std. Dev of  random effects (after)}
\label{fig5:sub19}
\end{subfigure}
~
\begin{subfigure}[t]{0.23\textwidth}
\includegraphics[width=\textwidth]{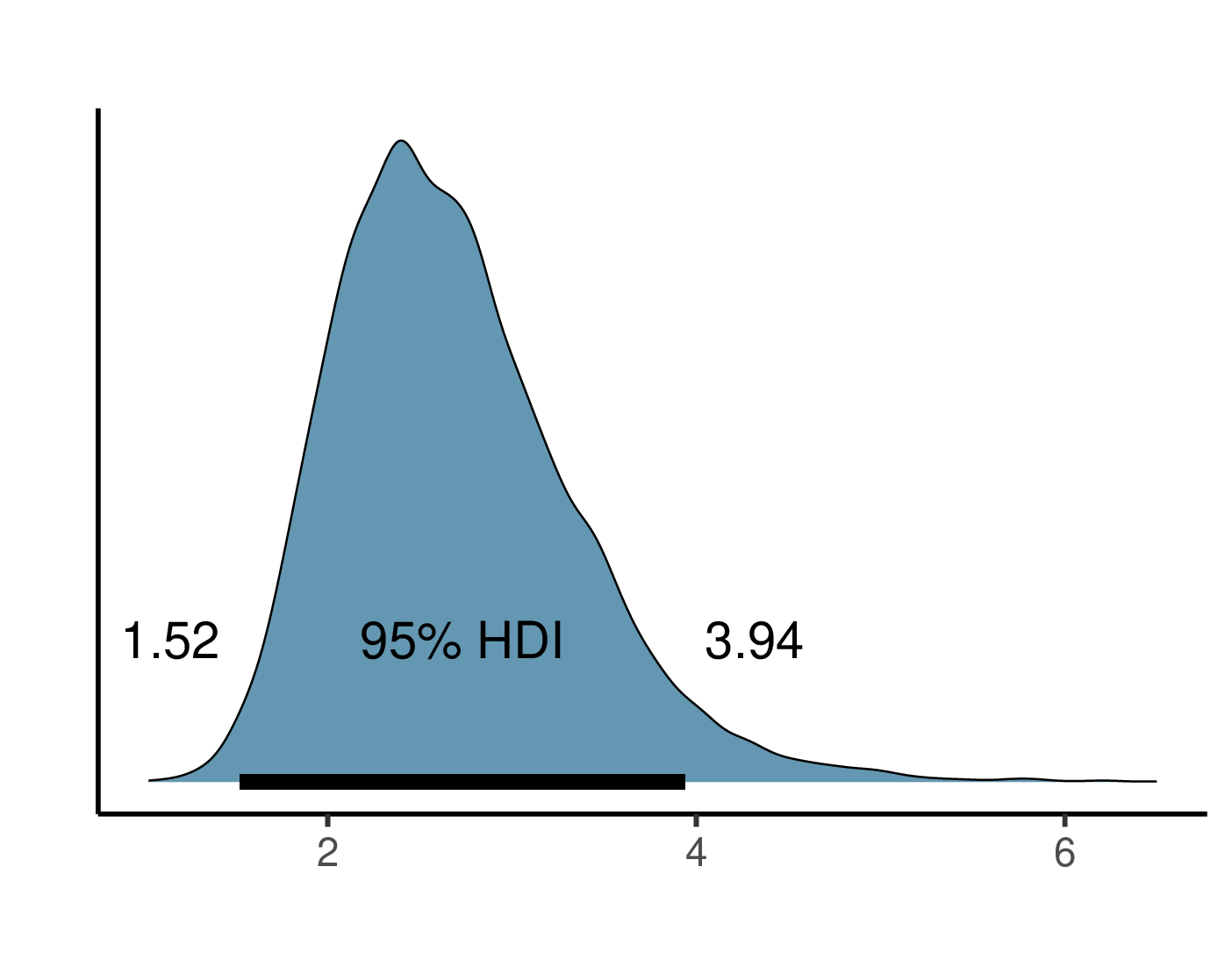}
\caption{Std. Dev of  random effects (no) }
\label{fig5:sub20}
\end{subfigure}
~
\caption{The density plot display the distribution of samples drawn from the posterior distribution for the fixed and random effects. The x-axis represents the range of possible values of the parameter of interest, which is the regression coefficient for the predictor variable. The y-axis shows the density of the posterior distribution for each coefficient value. The density represents the relative frequency of the coefficient occurring in the posterior distribution. Each plot is annotated with the highest posterior density interval (horizontal at the bottom). The width of the HDI reflects the uncertainty in the parameter estimate.}  
\label{fig5:allsubfigs}
\end{figure*}

Figure \ref{fig5:allsubfigs} displays the posterior distribution for the fixed effects and the random effects parameters. These plots display the probability distribution of a parameter given the observed data and prior information. The plots provide a visual representation of uncertainty in the parameter estimates (see table \ref{table:group-level-effects} and \ref{table:population-level-effects}), central tendency (e.g., mean), and spread. The uncertainty in the estimates is characterized by the width of the highest posterior density interval (HDI). The HDI represents a range of values for the parameter, covering a certain percentage of the posterior distribution. These values indicate the most credible points within this distribution.  A wider HDI indicates greater uncertainty, while a narrower HDI indicates more precision in the estimate \autocite{kruschke_bayesian_2013}. If the HDI includes zero, it indicates that the true value of the parameter might not be significantly different from zero, given the data and the model \autocite{kruschke_bayesian_2013}. The following sections provide a comprehensive analysis of the results.

\subsection{Random Effects }
Table \ref{table:group-level-effects} shows model results for random effects for the intercept  Driver ID. A random intercept is an extension of the fixed intercept in a regression model that allows the intercept to vary across different levels of a grouping factor. In this case, we have a random intercept for each driver (Driver ID). This means that each driver has their baseline probability of the response variable (Signal) for the after and no categories after accounting for the fixed effects. The random intercepts capture the variation in the baseline probabilities across drivers, which is not explained by the fixed effects (predictors). 

\begin{table}[H]
    \centering
    \topcaption[table:group-level-effects]{Random effects for the hierarchical model.}
    \begin{tabular}{lrrrrrrr}
    \hline
    Random effects & Estimate & Est.Error & l-95\% HDI & u-95\% HDI & Rhat & Bulk\_ESS & Tail\_ESS \\ \hline
        \multicolumn{8}{l}{\textbf{Standard deviation}} \\
        \multicolumn{8}{l}{\textbf{\textit{After}}} \\
        Intercept & 0.36 & 0.11 & 0.13 & 0.57 & 1.00 & 2554 & 2557 \\
        \multicolumn{8}{l}{\textbf{\textit{No}}} \\
        Intercept & 2.69 & 0.55 & 1.58 & 4.19 & 1.00 & 2685 & 4289 \\ \hline
    \end{tabular}
\end{table}

The column ``Estimate'' represents the variability in the intercepts and slopes among different levels of the grouping factor (Driver ID in this case). The ``Est.Error'' measures the uncertainty or variability in the estimate. A larger standard error indicates a greater degree of uncertainty in the estimate. The ``l-95\%'' HDI and ``u-95\% HDI'' is 95\%, the range within which we can be 95\% confident that the true parameter value lies. The ``Rhat'' is a convergence diagnostic used to assess whether the MCMC chains have converged to the target distribution. A Rhat value close to 1 indicates good convergence, and a value less than or equal to 1.1 is considered acceptable. Bulk ESS and Tail ESS  represent the effective sample size (ESS) measures. The ESS estimates the number of independent draws from the target distribution to which the MCMC sample is equivalent. A higher ESS indicates that the MCMC sample provides more information about the target distribution. 

The standard deviation of the random intercept for the level ``after'' suggests some variability in the ``after'' logits across different drivers. In other words, the likelihood of using a turn signal after starting a lane change varies across drivers when considering the average effect of all other factors.  The standard deviation of the random intercepts for the ``no'' level suggests substantial variability in the ``no'' logits across different drivers. In other words, the likelihood of not using a turn signal while changing lanes varies across drivers when considering the average effect of all other factors.

\subsection{Fixed Effects }

In table, \ref{table:population-level-effects}, estimates for the fixed effects in the ``after'' and ``no'' compared to the response level ``before'' of the response variable are presented. The sign of the coefficient of a predictor determines its relation with the outcome variable.  For a continuous predictor, the estimates represent the change in log odds associated with a one-unit increase in the respective predictor, holding all other predictors constant. For categorical predictors, the estimates represent the change in log odds associated with a change from the reference category to the respective category, holding all other predictors constant. 

\begin{table*}[]
\centering
\begin{threeparttable}
\topcaption[table:population-level-effects]{Fixed effects for the hierarchical model}
\begin{tabular}{lrrrrrrr}
\hline
Predictors & Estimate & Est.Error & l-95\% HDI & u-95\% HDI & Rhat & Bulk\_ESS & Tail\_ESS \\ \hline
\textit{\textbf{After}} & & & & & & & \\
Intercept & -1.67 & 0.51 & -2.68 & -0.7 & 1.00 & 12851 & 8654 \\
Speed & 0.02 & 0.01 & 0.01 & 0.04 & 1.00 & 12714 & 8803 \\
Direction (Right)\textsuperscript{1} & -0.81 & 0.11 & -1.03 & -0.59 & 1.00 & 17352 & 7662 \\
Rear Vehicle (Yes)\textsuperscript{2}  & -0.37 & 0.18 & -0.74 & -0.02 & 1.00 & 9408 & 7662 \\
Rear Gap & 0.19 & 0.09 & 0.02 & 0.37 & 1.00 & 9111 & 7480 \\
Lag Vehicle (Yes)\textsuperscript{2}  & -0.23 & 0.18 & -0.59 & 0.13 & 1.00 & 9104 & 7691 \\
Lag Gap & 0.07 & 0.09 & -0.12 & 0.24 & 1.00 & 9031 & 7608 \\
Traffic Density & -0.02 & 0.01 & -0.04 & 0.00 & 1.00 & 15617 & 7884 \\
Driver Type (Pro) \textsuperscript{3} & 0.30 & 0.18 & -0.04 & 0.65 & 1.00 & 5447 & 6251 \\
\textit{\textbf{No}} & & & & & & & \\
Intercept & -5.92 & 1.34 & -8.60 & -3.41 & 1.00 & 5553 & 5741 \\
Speed & 0.01 & 0.01 & -0.01 & 0.03 & 1.00 & 11882 & 7803 \\
Direction (Right)\textsuperscript{1}  & 0.02 & 0.22 & -0.40 & 0.47 & 1.00 & 15624 & 7487 \\
Rear Vehicle (Yes)\textsuperscript{2}  & -0.44 & 0.38 & -1.16 & 0.32 & 1.00 & 9182 & 7773 \\
Lag Vehicle (Yes)\textsuperscript{2}  & -1.10 & 0.35 & -1.83 & -0.46 & 1.00 & 8098 & 7838 \\
Lag Gap & 0.48 & 0.14 & 0.20 & 0.75 & 1.00 & 7283 & 7006 \\
Traffic Density & -0.07 & 0.03 & -0.12 & -0.02 & 1.00 & 16416 & 7826 \\
Driver Type (Pro) \textsuperscript{3} & 2.87 & 1.19 & 0.69 & 5.34 & 1.00 & 2647 & 4003 \\ \hline
\end{tabular}
\begin{tablenotes}
\small
\item \textsuperscript{1} Left is the reference level
\item \textsuperscript{2} No is the reference level
\item \textsuperscript{3} Non-Pro is the reference level
\end{tablenotes}
\end{threeparttable}
\end{table*}

The speed of the lane-changing vehicle is positively associated with the outcome variable level ``after''. This suggests that the probability of using a turn signal after starting a  lane change slightly increases with the per unit (km/h) increase in the speed of the subject vehicle.  The speed of the lane-changing  vehicle is also positively associated with the outcome variable level ``no''. This indicates an increased probability of not using at all before starting a lane change with a per unit increase in the speed of a lane-changing vehicle. However, the HDI for the level  ``no'' includes a zero indicating that the value of the parameter might not be significantly different from zero.  Overall these results indicate that high speeds are maybe the potential motivation behind not using the turn signal properly while changing lanes. 

The direction of lane change is negatively associated with the outcome variable level ``after''. This indicates the probability of using a turn signal after starting a lane change decreases while making a lane change to the right compared to the left. This could be because rightward lane changes frequently occur after overtaking slower vehicles and returning to the initial lane, or because rightward lane changes (in right-hand traffic) are often mandatory. In both cases, drivers may be more likely to use turn signals correctly to inform surrounding traffic. Conversely, the direction of the lane change has a positive association with the ``no'' outcome variable level, which seems to contradict the findings for the ``after'' level. However, the HDI includes zero, indicating that the parameter estimate may not be significantly different from zero. The discrepancy in the number of observations for each outcome variable level (after and no) could also contribute to the contradictory findings and uncertainty regarding the impact of lane change direction on the ``no'' outcome variable level. 

Having a rear vehicle in the current lane is negatively associated with the outcome variable level ``after''. This indicates that the probability of using a turn signal after starting a lane change decreases when a rear vehicle is in the current lane. However, a rear gap is positively associated with the outcome variable level ``after''. This indicates that even if there is a rear vehicle, the probability of using a turn signal after starting a lane change increase with one unit increase in the rear gap. This is generally the case in real traffic where surrounding vehicles affect the lane change behaviour \autocite[e.g. as described in][]{moridpour2010effect}. The association between the rear vehicle and the rear gap with outcome variable level ``no'' is the same as level ``after''.  However, the HDI for both predictors includes zero, indicating that the parameter estimate may not be significantly different from zero.

The presence of a lag vehicle shows a negative association with both ``after'' and ``no'' outcome variable levels, whereas the lag gap shows a positive association with these levels. Although these results align with those of rear vehicle and rear gap, the inclusion of zero in the confidence interval suggests uncertainty about the true impact on the outcome variable. The presence of rear and lag vehicles suggests increased overall traffic density. In our study, traffic density refers to the number of vehicles per kilometre per lane. The findings show a negative association between traffic density and both ``after'' and ``no'' outcome variable levels, suggesting that a one-unit increase in traffic density reduces the probability of turn signal usage being  as ``after'' and "no." However, the inclusion of zero in the HDI for the ``after'' level indicates that the parameter estimate may not be significantly different from zero.

Last but not least, driver type (Pro) was positively associated with both ``after'' and ``no'' outcome variable levels. This suggests that the probability of using a turn signal after starting a lane change and not using a turn signal increases for professional drivers. However, the inclusion of zero in HDI for the ``after'' level and a very wide HDI for the ``no'' level indicates uncertainty about the true effect on the outcome variable. Furthermore, the discrepancy in the number of observations in the ``no'' level for the driver type could also contribute to these findings. Therefore, these findings should be interpreted with caution. 

\section{General Discussion and Conclusion}\label{discussion}
This research examined the use of turn signals by Swedish (European) drivers during lane changes. Data for this study was collected on actual roads in Gothenburg, Sweden, observing drivers' natural behaviour without interference from the experimenter. Out of the 103 participants, the majority were  non-professional drivers (everyday drivers), while approximately 10\% were professional test drivers.  The dataset included information on turn signal usage, the speed of the vehicle making a lane change, the direction in which the lane change occurred, the presence of and gaps between surrounding vehicles, as well as the overall traffic density. The turn signal usage was categorized into three levels: before starting a lane change, after starting a lane change, and not using the turn signal at all. The "before" level was selected as the baseline level due to its higher frequency and overall significance. 

Our results showed that the turn signal usage in this study was very high, with approximately 93\% of cases involving the use of turn signals. The results indicate higher compliance with turn signal usage than what has been reported in previous studies \autocite[e.g.][]{lee_comprehensive_2004, ponziani_turn_2012, wang2014investigation, wang2019analysis, lin2019effect}. There may be several reasons for this high level of compliance. One possible reason could be that Swedish drivers have a higher compliance rate and a positive attitude towards adhering to traffic rules, as reported in previous studies \autocite{warner2009cross, sinclair2013attitudes}. However, a substantial number (About 33\%) of lane changes were observed where turn signals were used improperly. While there is no universally accepted definition of proper turn signal use in literature, it is generally understood to mean activating the turn signal before starting a lane change. Furthermore, in about 7\% of the cases, a turn signal was not used at all. Overall, our results showed that only about 60\% of the lane changes started with proper turn signal usage.  

We further examined factors that impact turn signal usage through the application of Bayesian hierarchical modelling. The model consisted of fixed effects for variables such as speed, direction, rear vehicle, rear gap, lag vehicle, lag gap, traffic density, and driver type. Additionally, it included a random effect for Driver ID. The random effect for the Driver ID allowed to capture the variability in turn signal usage between different drivers.   The results show  that a range of factors influences turn signal usage, each with distinct levels of uncertainty. Highlighting the uncertainty in the findings not only acknowledges the fundamental principles of Bayesian analysis but also promotes a nuanced comprehension of the relationships and insights obtained from the data \autocite{kruschke_bayesian_2013}. 


 The results for the random effects indicated that the probability of using a turn signal after initiating a lane change is not consistent across drivers when considering the average effect of all other factors included in the model. Similarly, the probability of not using a turn signal while changing lanes is also inconsistent across drivers when considering the average effect of all other factors in the model. This variability suggests that individual driver behaviour and preferences may play a role in the likelihood of using a turn signal during lane change. Future research should investigate the  factors contributing to individual differences in turn signal usage during lane changes. These factors may include but are not limited to,  driving habits, personality traits, or other unmeasured factors.  

The fixed effects in the model capture the overall relationship between the predictors and the response variable, representing the average effects of each predictor on the response variable across all drivers. The speed of the lane-changing vehicle was positively associated with both levels of the outcome variable. That means, with an increase in speed, the probability of either not using a turn signal or using it after starting a lane change also increased. A plausible reason for this could be that drivers tend to be less worried about vehicles in the intended lane when driving at faster speeds. 

The direction in which a vehicle moves during a lane change can affect the overall manoeuvre. The direction of the lane change is typically associated with the underlying motivation for making the change. For example, drivers may choose to make discretionary lane changes to the left (in a right-hand traffic scenario) to gain a speed advantage. Lane changes to the right typically occur due to mandatory reasons, such as taking an exit. The motivation behind a lane change subsequently influences the characteristics of the manoeuvre \autocite{wang2019analysis}. Our findings indicate that when drivers make lane changes to the right, they are less likely to use their turn signal after initiating the lane change. Conversely, the direction of the lane change has a positive association with the  outcome variable level ``no''. This observation is consistent with \autocite{lee_comprehensive_2004}, which reported that during right lane changes in the US, drivers might not deem it necessary to signal their intentions after passing a slower lead vehicle. Please note that the results for the outcome variable ``no'' should be interpreted cautiously due to the inclusion of zero in the HDI.

According to prior research, the presence of vehicles in the current and target lane impacts lane change behaviour \autocite{moridpour2010effect}. A slow vehicle in the current lane may prompt drivers to switch lanes in order to increase their speed. Similarly, the rear vehicle in the current lane and lag and lead vehicle in the target lane have an impact on overall lane change (e.g., gap acceptance or lane change duration). Similarly, the presence of a rear vehicle in the current lane, as well as lag and lead vehicles in the target lane, can influence the overall lane change process (e.g., gap acceptance or lane change duration). Our findings indicate that the probability of using a turn signal after starting a lane change, or not using it at all, decreases when there are rear and lag vehicles present. This is also reflected by the traffic density, which suggests that an increase in traffic density reduces the  probability of using a turn signal after starting a lane change or not using it at all.  However, as the rear gap and lag increase, the probability of using a turn signal after starting a lane change or not using it at all also increases. In other words. the impact of surrounding vehicles  decreases with larger gap sizes. Future research should explore the specific gap at which the influence of the gap on turn signal usage during lane changes becomes negligible.

In our study, the main difference between driver types was their training and driving experience. Professional drivers were trained to operate prototype vehicles for testing specific features, while they  had more experience behind the wheel than non-professional drivers.  However, due to the imbalanced data and the discrepancy in the number of observations for each outcome variable level, the results concerning the driver type should be interpreted cautiously. Although a comparison between professional and non-professional drivers was not a focus of this study, we also performed individual BHM analyses of the two datasets independently. The comparison results are provided in appendix \ref{appendixA}. The results are provided in terms of posterior distribution density plots for both random and fixed effects. 

To sum up, the results of this study make a valuable contribution to the current body of literature by offering in-depth insights into the usage of turn signals during lane changes. Using turn signals while changing lanes has important effects on traffic safety and flow. Failing to use a turn signal while changing lanes can cause traffic disruption and increase the risk of collisions. Using turn signals as a direct means of communication is crucial for the future of mixed traffic, which includes both autonomous vehicles and traditional vehicles. Automated vehicles (AVs) are designed to follow strict rules, but they may face challenges in effectively communicating or negotiating right-of-way in certain situations\autocite{metz2018developing, farber2016communication}. In this regard, a human driver's lane change without using a turn signal can cause problems for an AV. For instance, an AV may not be able to signal its intentions or force the driver to cancel a lane change. In such cases, the AV's reaction would likely be a sudden deceleration, potentially causing disruptions to traffic and discomfort for its occupants and other vehicles on the road.

Finally, like any scientific study, this research also had some limitations. The data were collected only in a specific city in Sweden. Future studies should consider data from other parts of Europe (and other parts of the world) to gain a more comprehensive understanding of turn signal usage behaviour among European drivers. The current study only focused on passenger cars. Thus, future studies should also consider different vehicle types, including heavy vehicles, as they greatly impact traffic. We showed the presence of lag, and rear vehicle impacts the turn signal usage. As a result, it would be valuable for future studies to take into account the types of surrounding vehicles present. We also found that the direction of lane change impacts the usage of turn signals. The direction could also be related to the motivation behind the lane change (e.g., discretionary or mandatory lane change), which in turn has an impact on the overall lane change process \autocite{vechione2018comparisons}. Therefore, future studies should consider investigating the differences in turn signal usage based on motivation. 

\section*{Conflict of Interest}\label{sec6}
The authors declare that they have no known competing financial interests or personal relationships that could have appeared to influence the work reported in this paper.
\section*{Acknowledgment}\label{sec7}
The study was carried out as part of the SHAPE-IT project, receiving funding from the European Commission’s Horizon 2020 Framework Programme under the Marie Skłodowska-Curie Actions initiative (Grant agreement 860410). We thank Volvo Car Corporation and the L3Pilot (EC grant agreement: 723051) project for collecting the data and providing us access to it. We also want to thank SAFER Vehicle and Traffic Safety Center at Chalmers, Gothenburg, Sweden, for providing us with the necessary facilities to extract data.  
\section*{Declaration of Generative AI and AI-assisted technologies in the writing process}
During the preparation of this work the authors used GPT-4 (ChatGPT) and GrammarlyGO in order to improve readability of some sentences. After using this tool/service, the authors reviewed and edited the content as needed and take full responsibility for the content of the publication.
\printbibliography

\clearpage
\appendix
\section{Appendix A}
\renewcommand{\thefigure}{\Alph{section}.\arabic{figure}}
\setcounter{figure}{0}

\begin{figure*}[h!]
\centering
\begin{subfigure}[t]{0.5\textwidth}
\includegraphics[width=\textwidth]{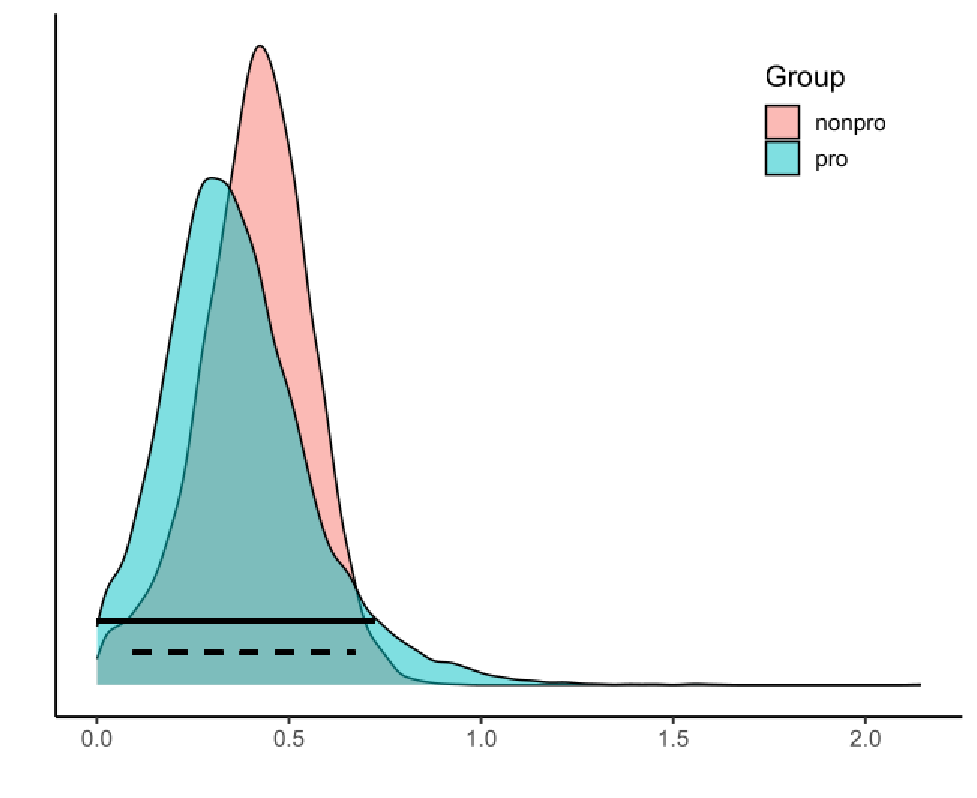}
\caption{Std. Dev of random effects
(after)}
\label{figA1:sub2}
\end{subfigure}%
~
\begin{subfigure}[t]{0.5\textwidth}
\includegraphics[width=\textwidth]{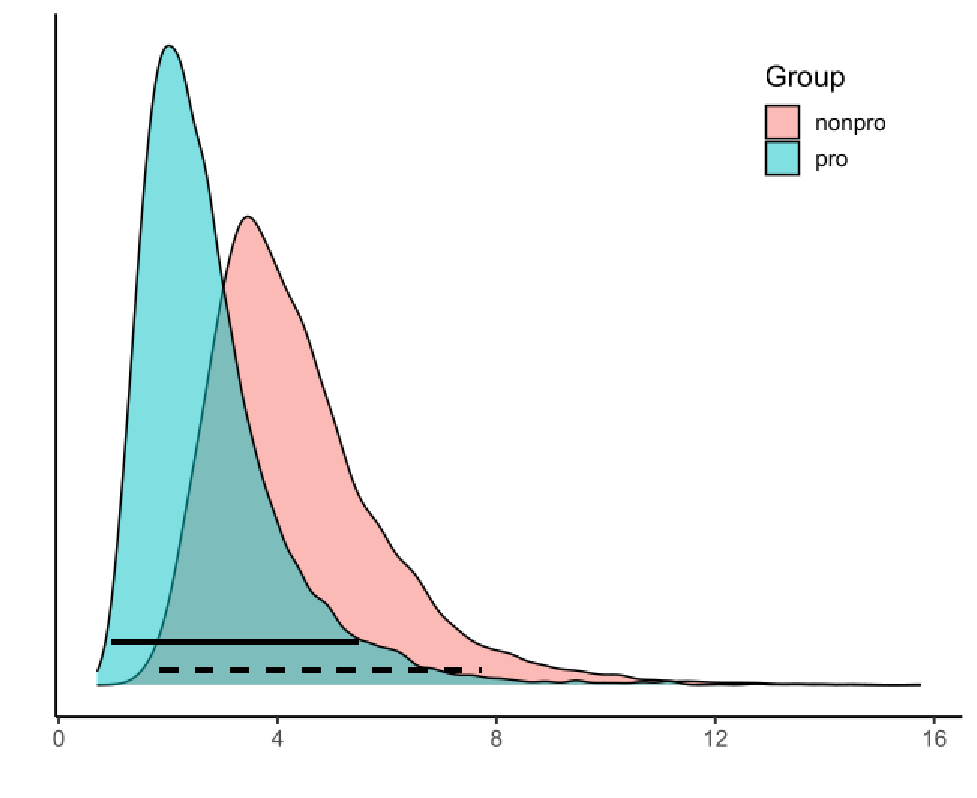}
\caption{Std. Dev of random effects
(no)}
\label{figA1:sub3}
\end{subfigure}
~
\caption{The density plot display the distribution of samples drawn from the posterior distribution for the random effects for both levels of outcome variable (i.e. after and no). The x-axis represents the range of possible values of the parameter of interest, which is the standard deviation. The y-axis shows the density of the posterior distribution for each coefficient value. The density represents the relative frequency of the coefficient occurring in the posterior distribution. Each plot is annotated with the highest posterior density interval (horizontal line at the bottom). The width of the HDI reflects the uncertainty in the parameter estimate.}  
\label{figA1:allsubfigs}
\end{figure*}

This section presents a comparison of the additional models that we ran. The priors and the parameters (e.g. a number of chains and iterations)  for each additional model were kept the same as in the original model. Please note that the number of observations and drivers in each dataset differs. This difference in sample size might affect the precision of the estimates and the width of the HDI intervals. Figure \ref{figA1:allsubfigs} shows the posterior distribution for the random effects for both levels of the outcome variable (i.e. after and no). The extent to which both plots overlap  can give us an idea of the difference between the two groups \autocite{kruschke_bayesian_2013}. For example, figure \ref{figA1:sub2} shows that overlap in posterior distribution for the outcome variable level ``after'' is lower than the level ``no'' in figure \ref{figA1:sub3}. In other words, professional drivers show less variability than non-professional drivers. However, further research is required considering a more balanced dataset.

\begin{figure*}[h!]
\centering
\begin{subfigure}[t]{0.25\textwidth}
\includegraphics[width=\textwidth]{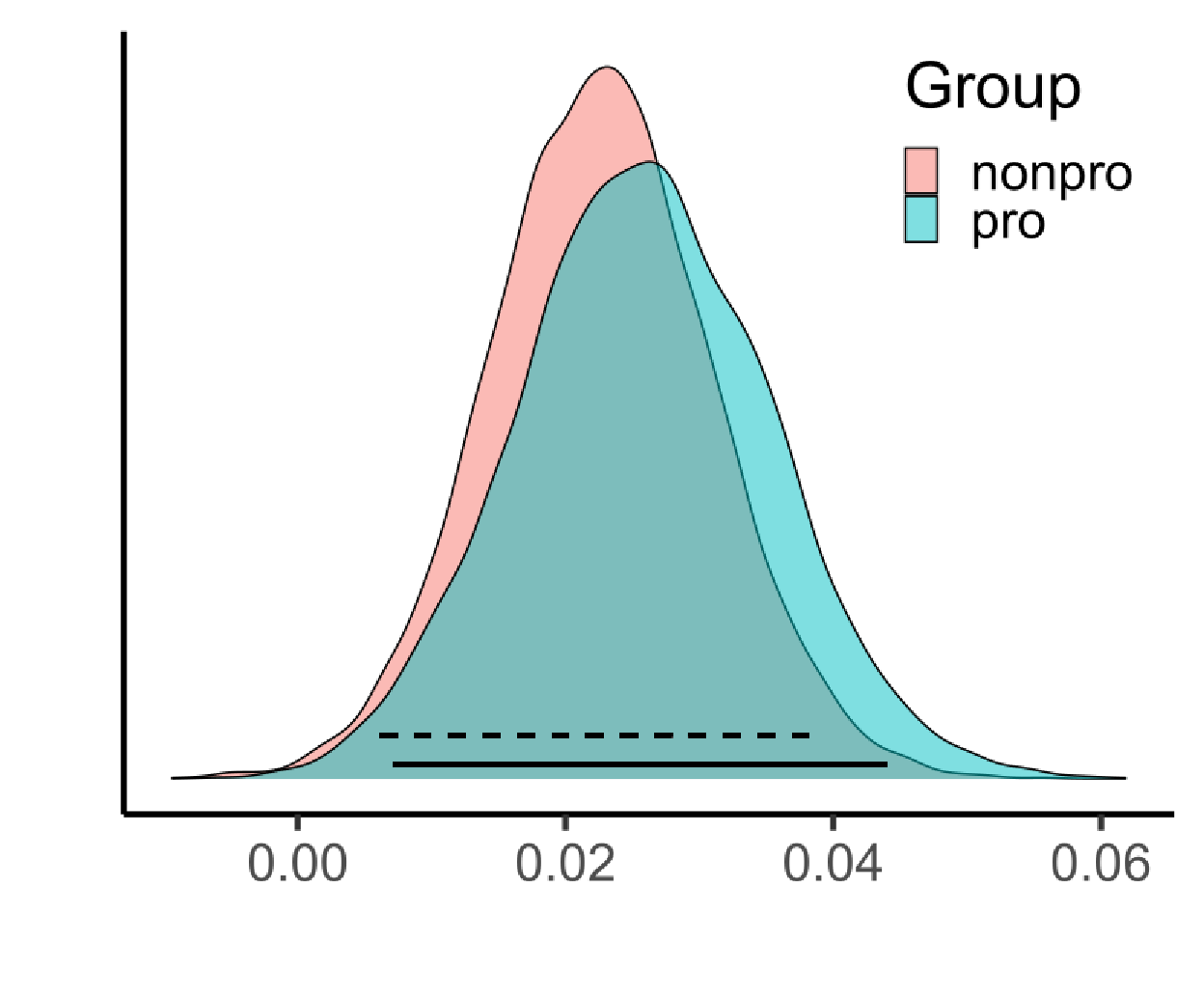}
\caption{Speed (after) }
\label{figA:sub2}
\end{subfigure}%
~
\begin{subfigure}[t]{0.25\textwidth}
\includegraphics[width=\textwidth]{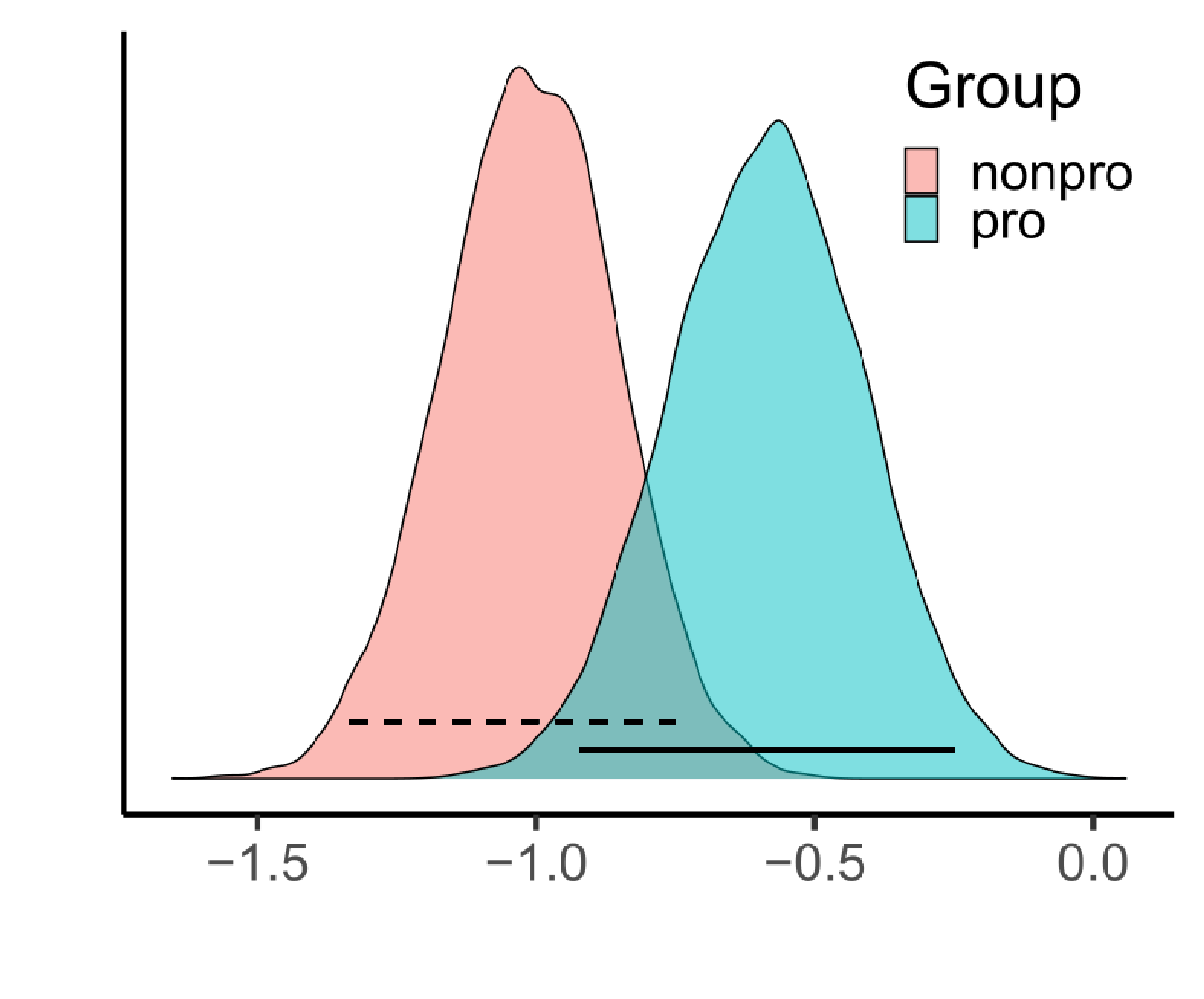}
\caption{Direction (after)}
\label{figA:sub3}
\end{subfigure}
~
\begin{subfigure}[t]{0.25\textwidth}
\includegraphics[width=\textwidth]{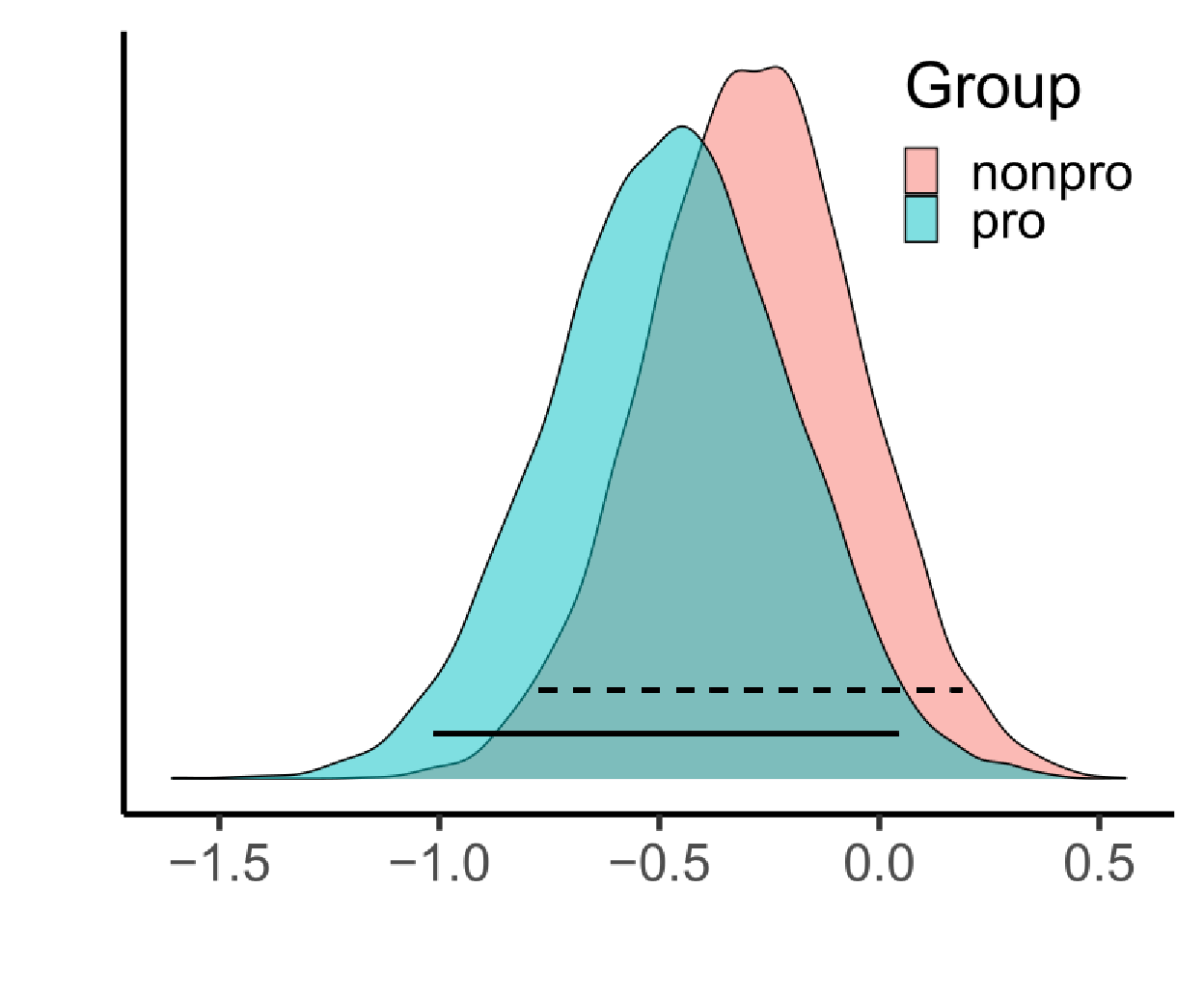}
\caption{Rear vehicle (after)}
\label{figA:sub4}
\end{subfigure}
~
\begin{subfigure}[t]{0.25\textwidth}
\includegraphics[width=\textwidth]{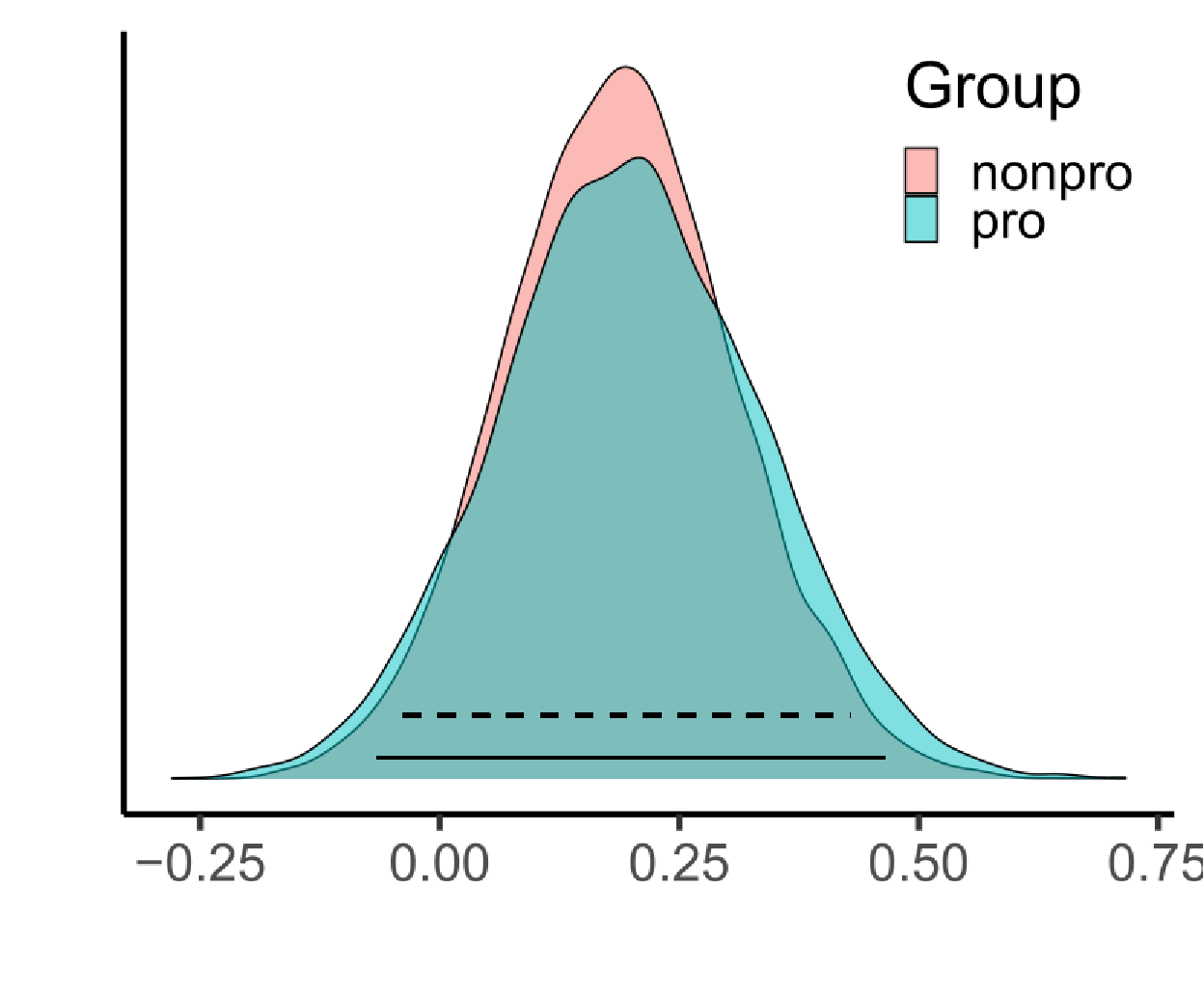}
\caption{Rear gap (after)}
\label{figA:sub5}
\end{subfigure}
~
\begin{subfigure}[t]{0.25\textwidth}
\includegraphics[width=\textwidth]{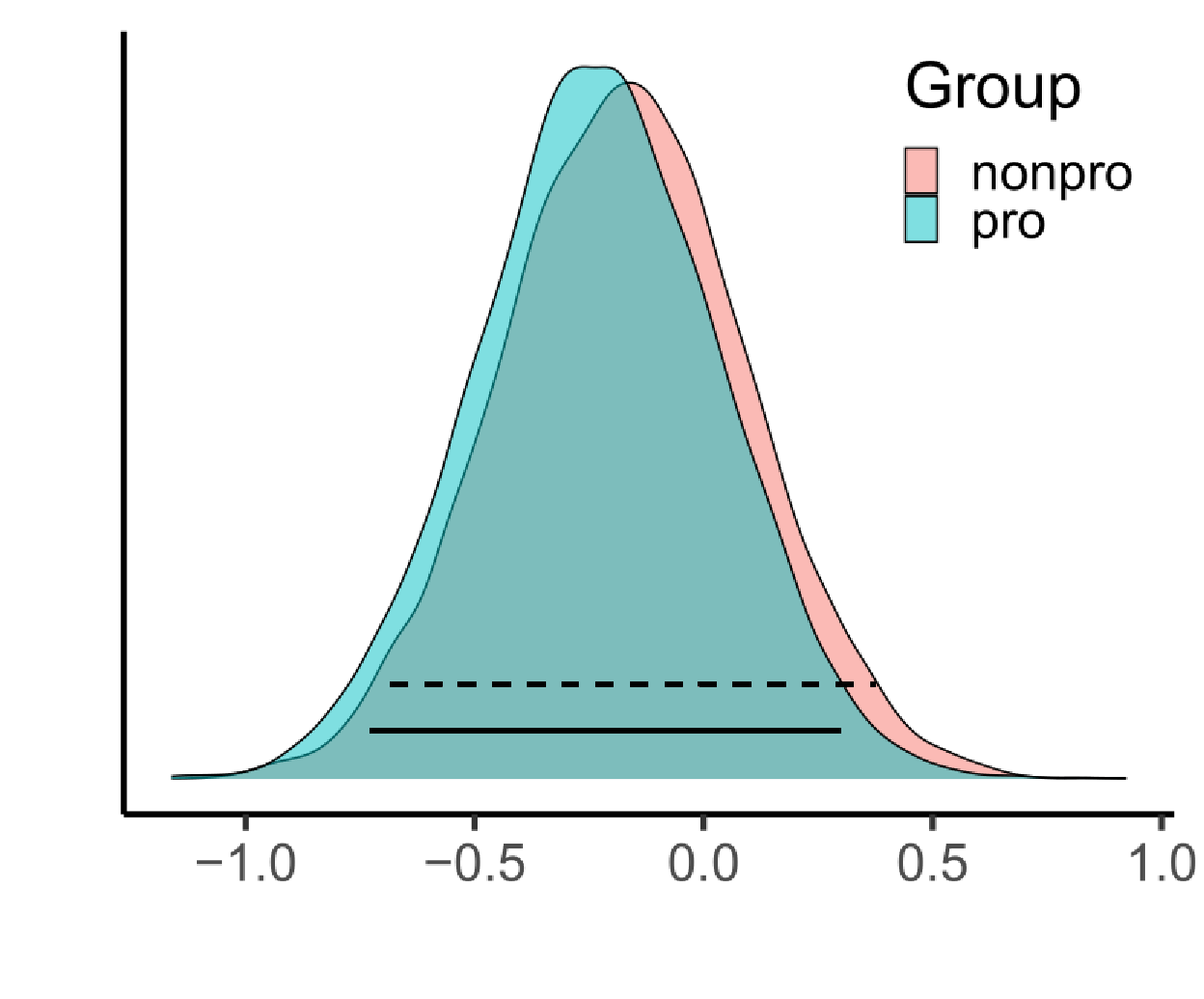}
\caption{Lag vehicle (after)}
\label{figA:sub6}
\end{subfigure}
~
\begin{subfigure}[t]{0.25\textwidth}
\includegraphics[width=\textwidth]{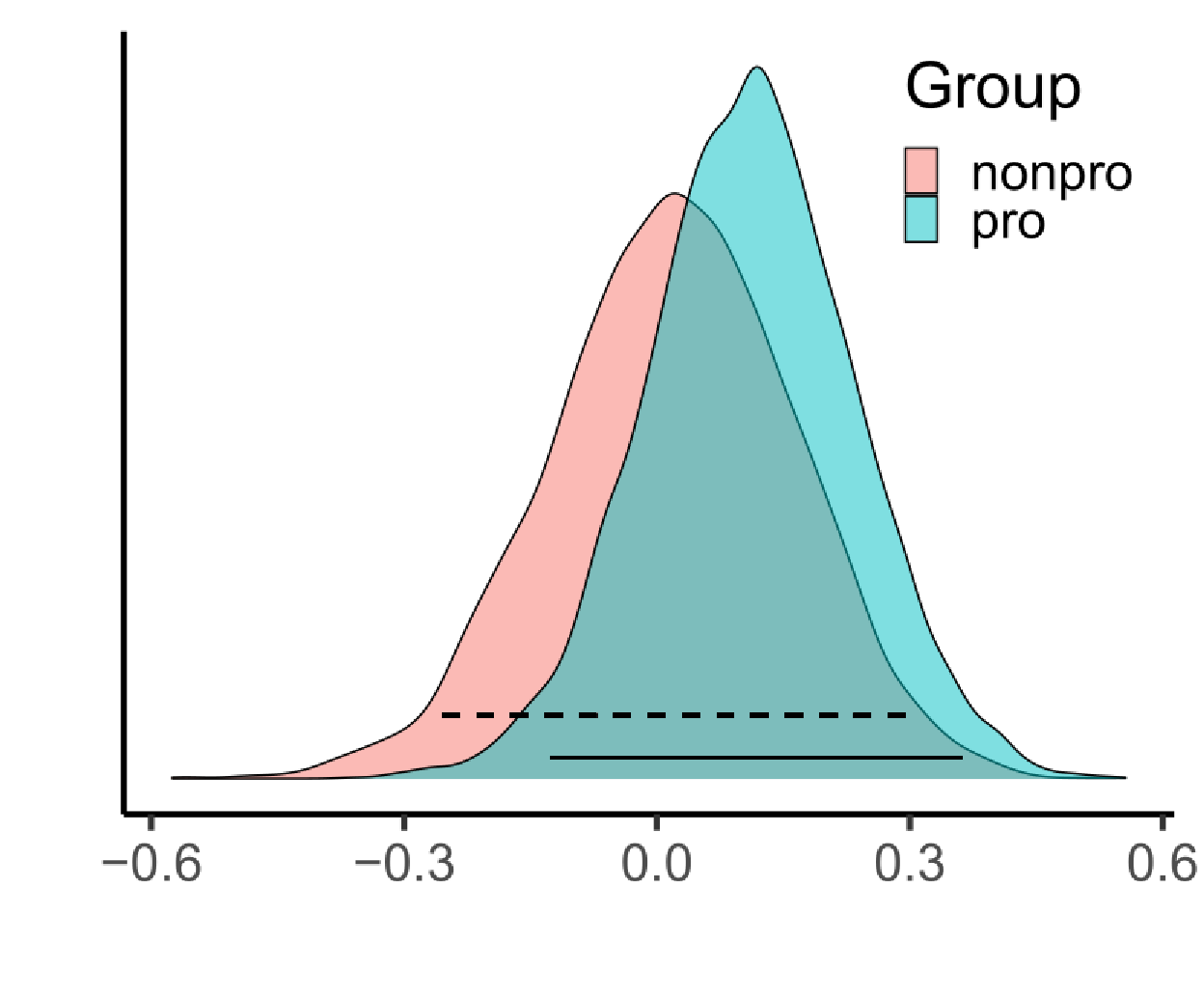}
\caption{Lag gap (after)}
\label{figA:sub7}
\end{subfigure}
~
\begin{subfigure}[t]{0.25\textwidth}
\includegraphics[width=\textwidth]{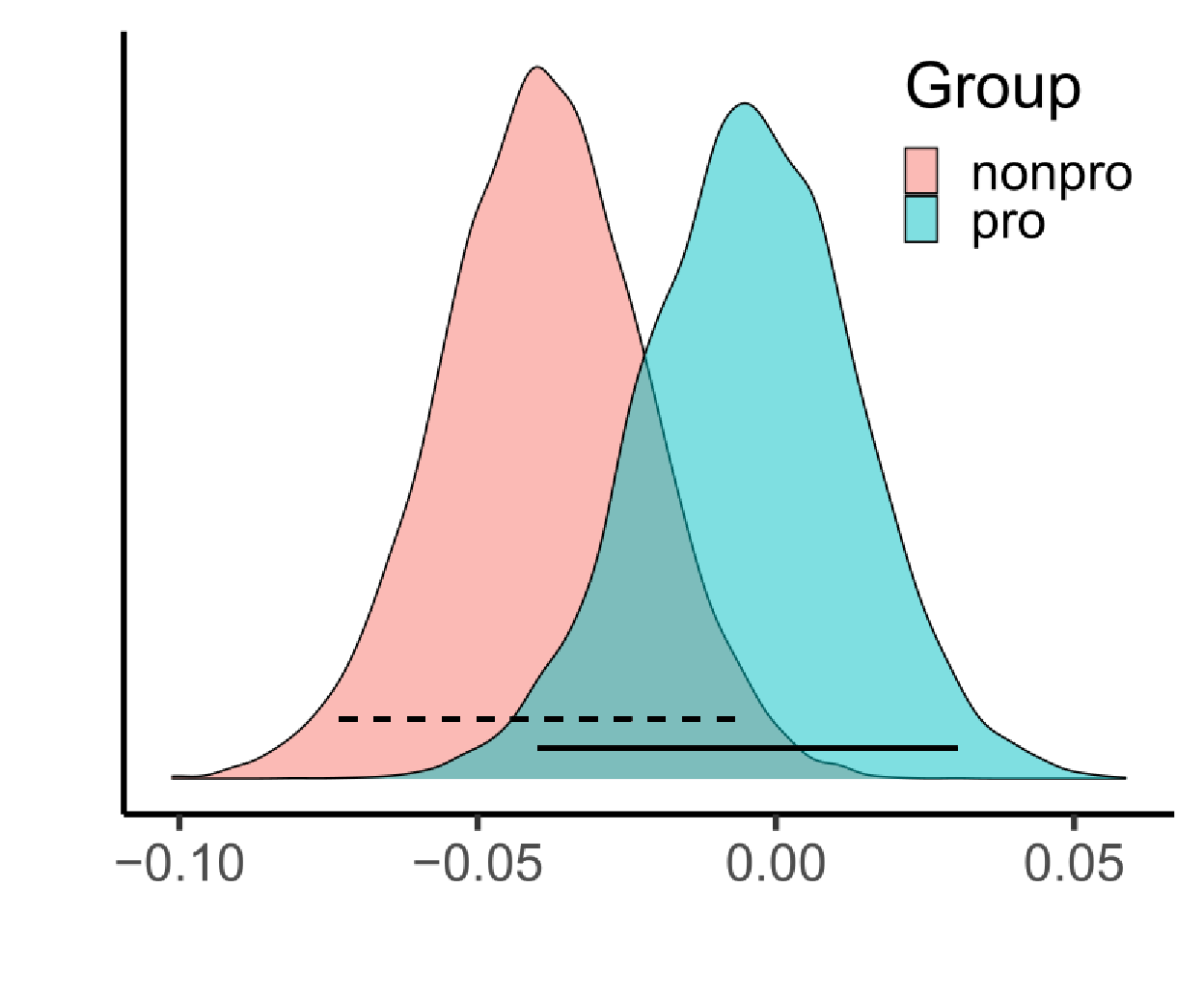}
\caption{Traffic density (after)}
\label{figA:sub8}
\end{subfigure}
~
\begin{subfigure}[t]{0.25\textwidth}
\includegraphics[width=\textwidth]{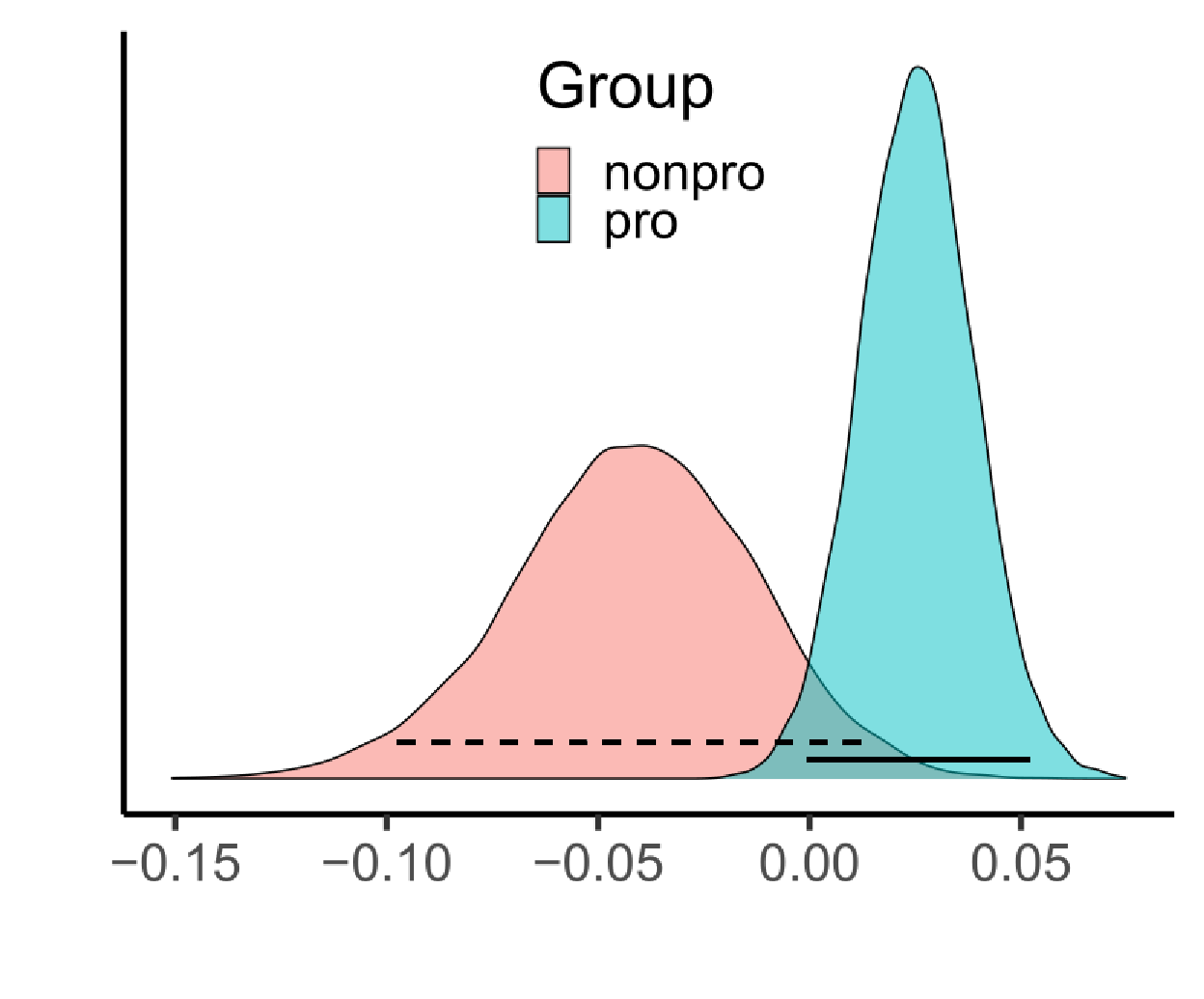}
\caption{Speed (no)}
\label{figA:sub11}
\end{subfigure}
~
\begin{subfigure}[t]{0.25\textwidth}
\includegraphics[width=\textwidth]{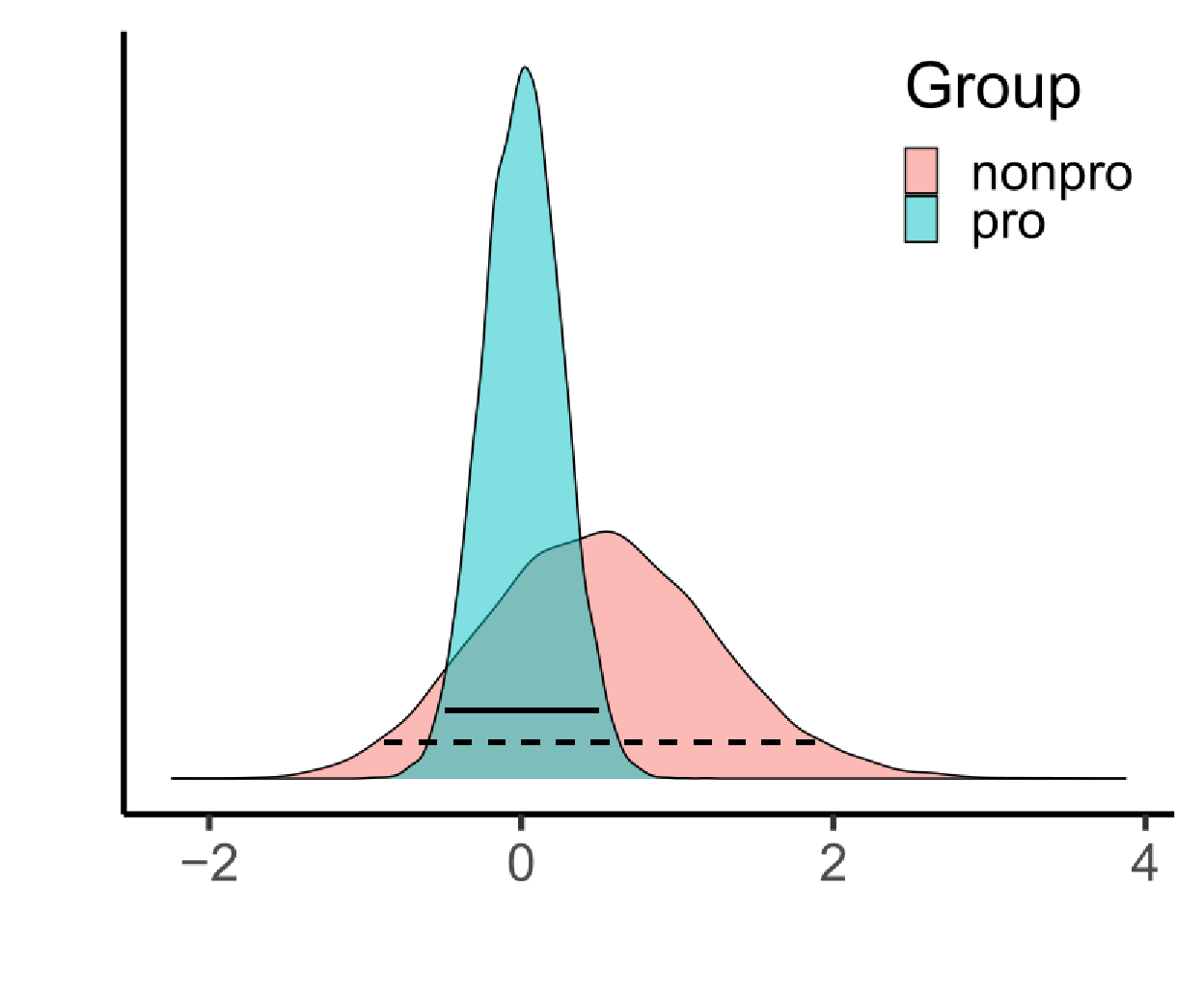}
\caption{Direction (no)}
\label{figA:sub12}
\end{subfigure}
~

\begin{subfigure}[t]{0.25\textwidth}
\includegraphics[width=\textwidth]{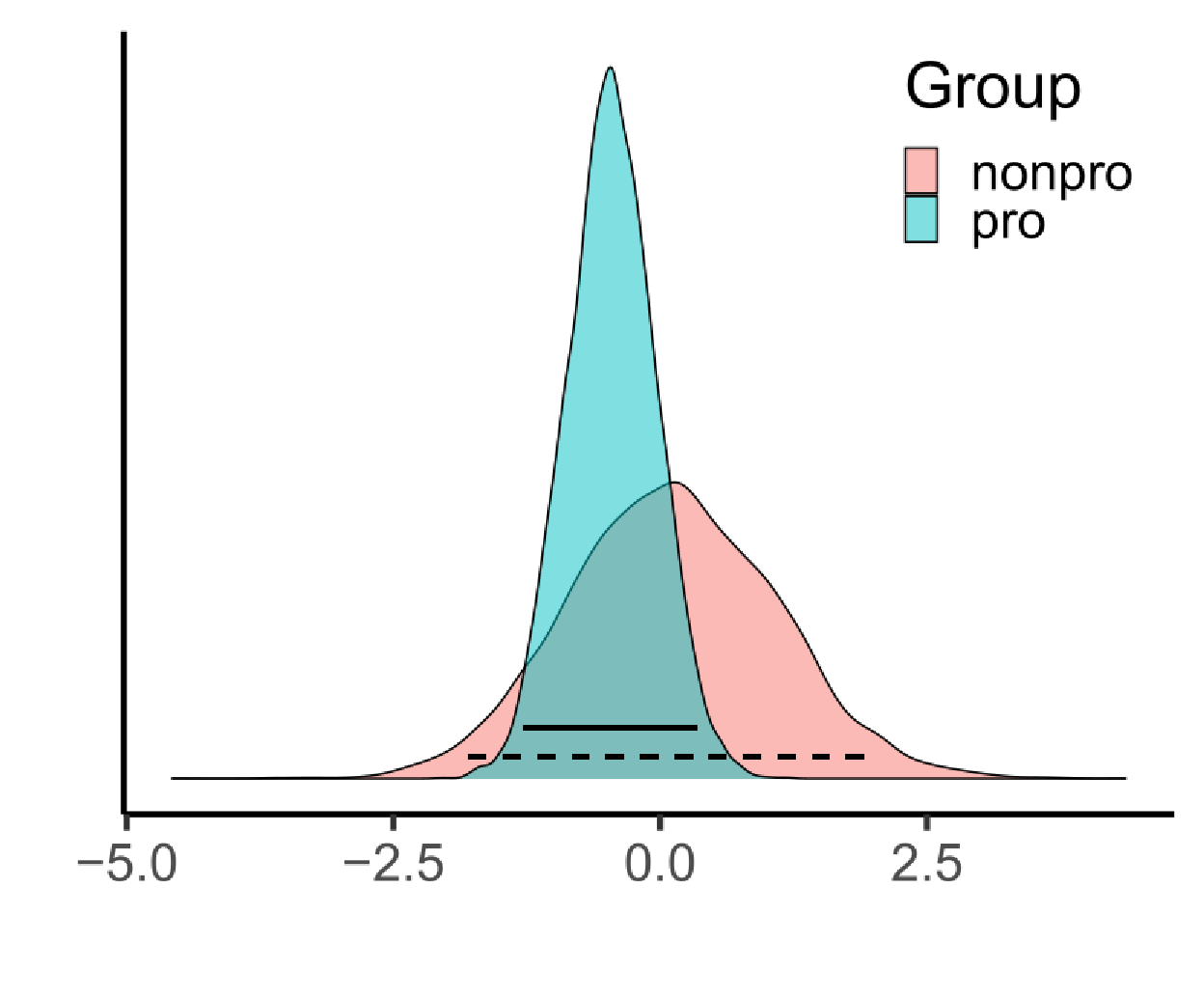}
\caption{Rear vehicle (no)}
\label{figA:sub13}
\end{subfigure}
~
\begin{subfigure}[t]{0.25\textwidth}
\includegraphics[width=\textwidth]{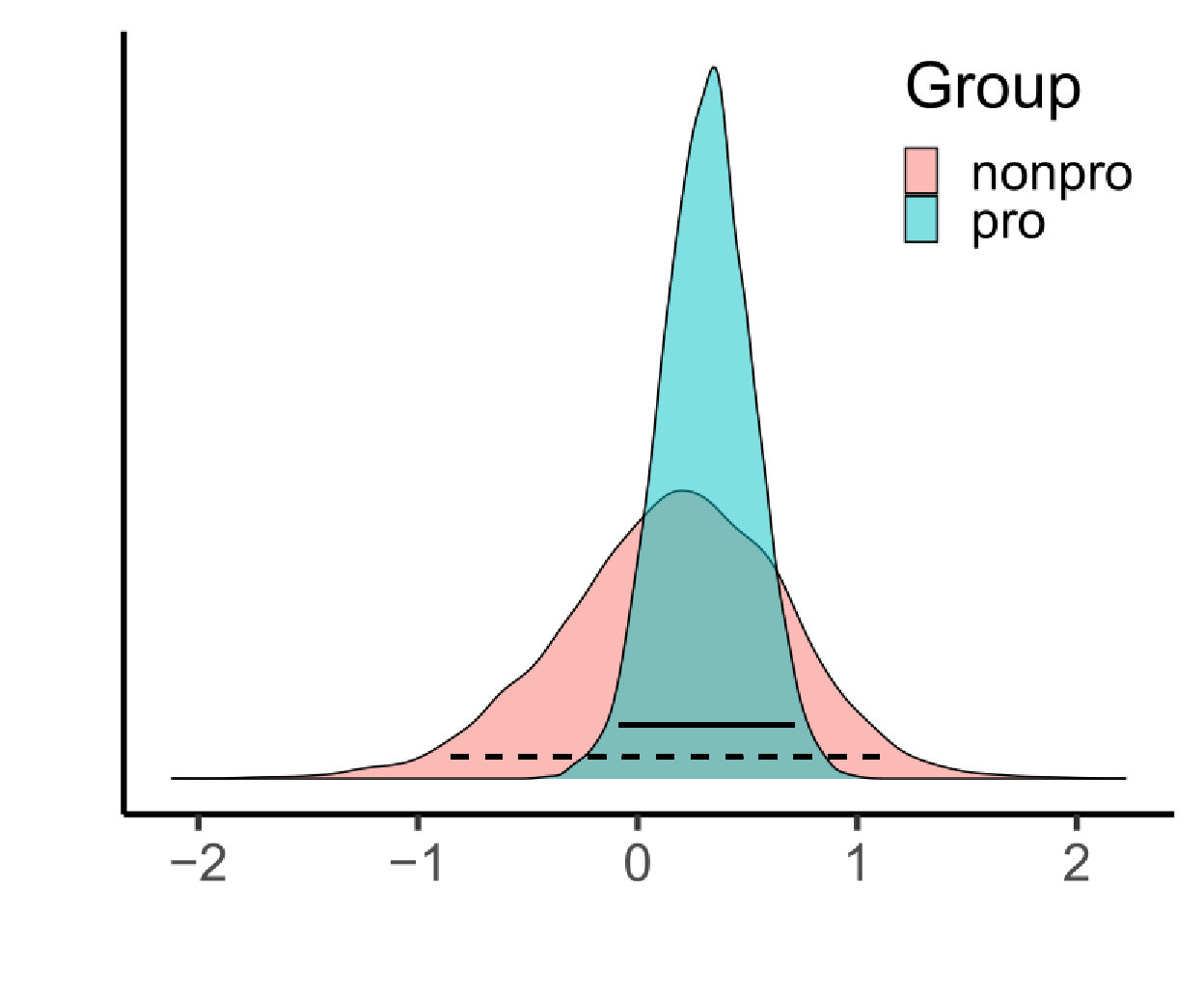}
\caption{Rear gap (no)}
\label{figA:sub14}
\end{subfigure}
~
\begin{subfigure}[t]{0.25\textwidth}
\includegraphics[width=\textwidth]{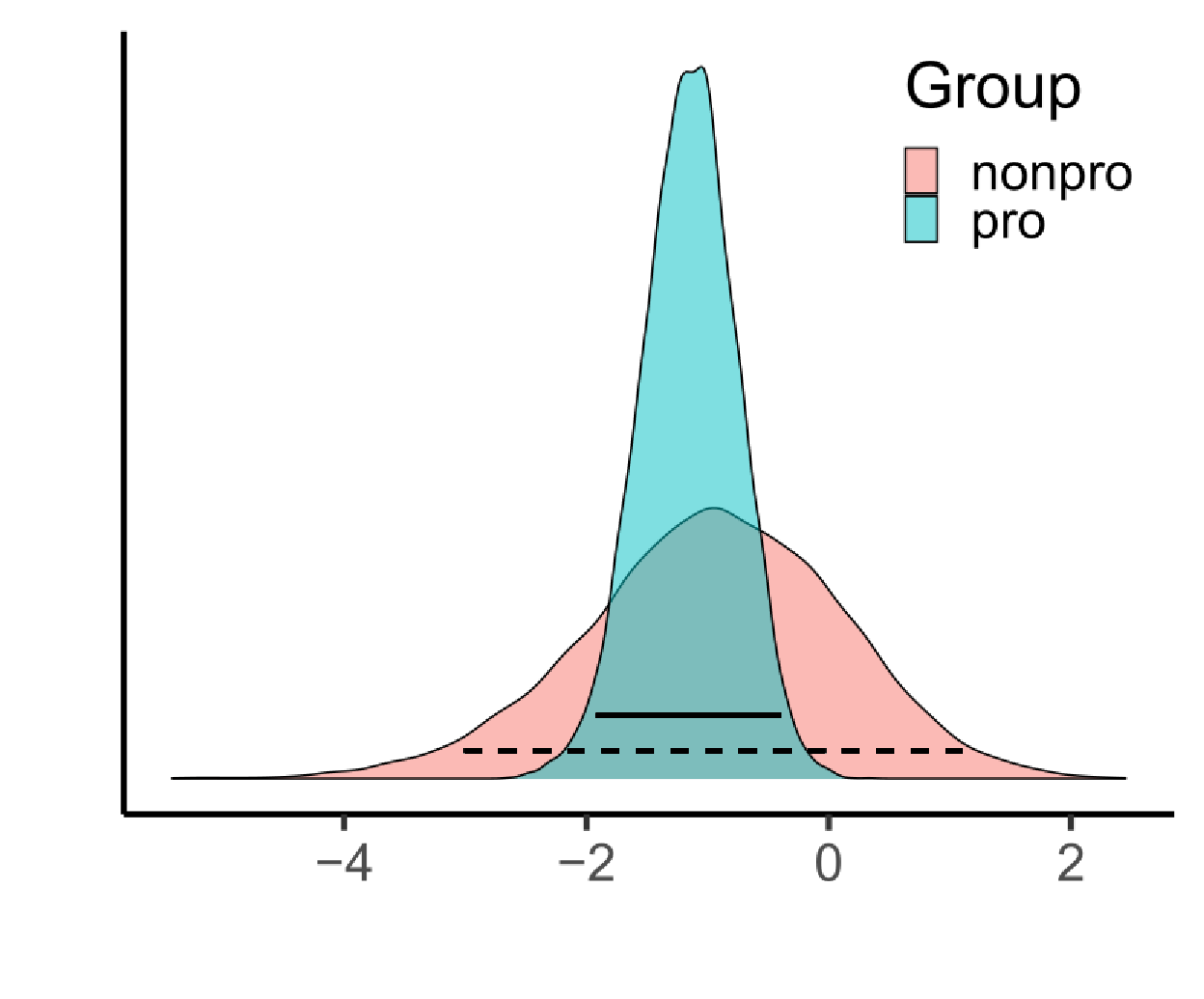}
\caption{Lag vehicle (no)}
\label{subfig:random_additional_model1}
\end{subfigure}
~
\begin{subfigure}[t]{0.25\textwidth}
\includegraphics[width=\textwidth]{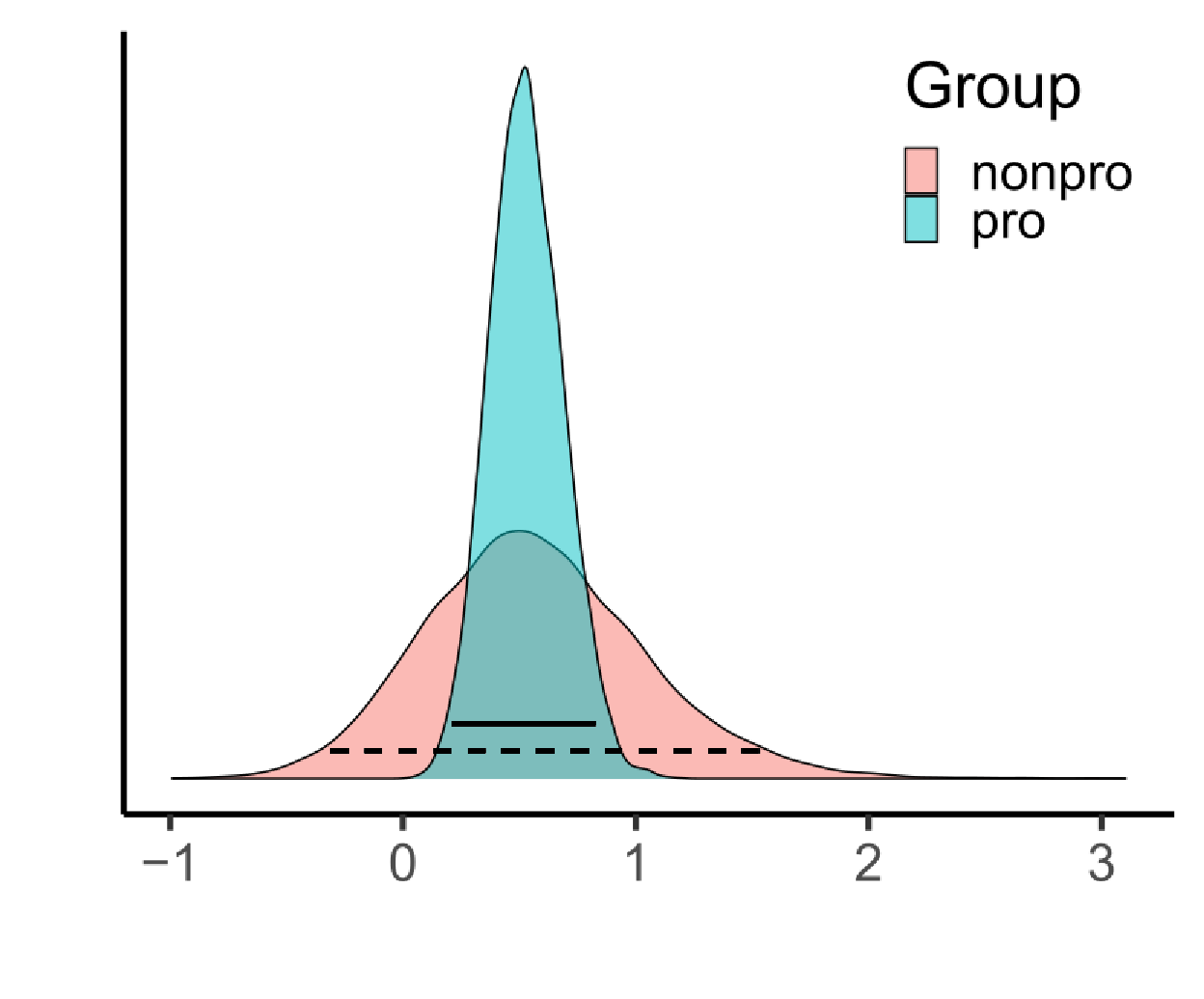}
\caption{Rear gap (no)}
\label{figA:sub16}
\end{subfigure}
~
\begin{subfigure}[t]{0.25\textwidth}
\includegraphics[width=\textwidth]{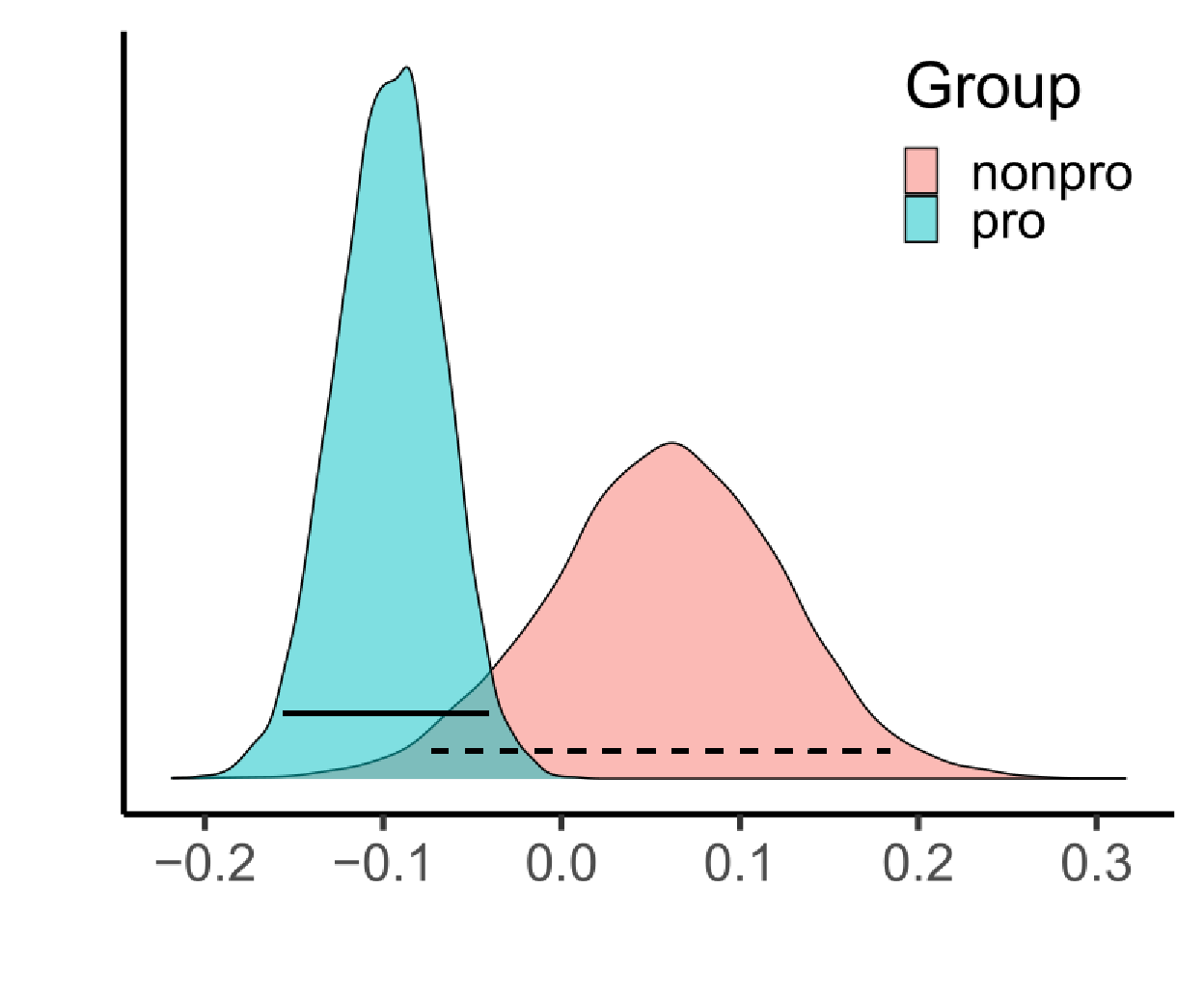}
\caption{Traffic density (no)}
\label{subfig:random_additional_model2}
\end{subfigure}
~
\caption{The density plot display the distribution of samples drawn from the posterior distribution for the fixed effects for both levels of outcome variable (i.e. after and no). The x-axis represents the range of possible values of the parameter of interest, which is the regression coefficient for the predictor variable. The y-axis shows the density of the posterior distribution for each coefficient value. The density represents the relative frequency of the coefficient occurring in the posterior distribution. Each plot is annotated with the highest posterior density interval (horizontal line at the bottom). The width of the HDI reflects the uncertainty in the parameter estimate.}  
\label{fig:random_additional_model}
\end{figure*}

Figure \ref{subfig:random_additional_model2} shows  the posterior distribution for the fixed effects for both levels of the outcome variable (i.e. after and no). For the level ``after'' we can see a substantial variability only for the direction of lane change and traffic density as shown in subfigure \ref{figA:sub3} and \ref{figA:sub8}. However, for the level ``no'' we can see a substantial variability in all the variables. This is because the all cases where a turn signal was not used belong to the professional driver group.


\end{document}